\documentclass[aps,showpacs,nofootinbib,amsmath,amssymb,preprint]{revtex4}
\usepackage{bm}
\usepackage{graphicx}
\usepackage{amsthm}

\begin{document}

\title{
Semi-analytic expressions for quasinormal modes of slowly rotating Kerr black holes
}

\author{
Yasuyuki Hatsuda and Masashi Kimura
}

\affiliation{
Department of Physics, Rikkyo University, Tokyo 171-8501, Japan
}

\date{\today}
\preprint{RUP-20-23}

\begin{abstract}
We provide semi-analytic expressions for quasinormal mode frequencies of slowly rotating Kerr black holes
to the quadratic order in the rotation parameter.
We apply the parametrized black hole quasinormal mode ringdown formalism to the Chandrasekhar-Detweiler equation and the Sasaki-Nakamura equation instead of the Teukolsky equation, and compare our result with the previous numerical calculations.
\end{abstract}

\maketitle

\section{Introduction}

The parametrized black hole quasinormal mode (QNM) ringdown formalism~\cite{Cardoso:2019mqo, McManus:2019ulj, Kimura:2020mrh}
is an efficient way to calculate QNM frequencies
when a master equation of the gravitational perturbation is slightly deviated
from that in the Schwarzschild case, {\it i.e.,} the Regge-Wheeler or Zerilli equation~\cite{Regge:1957td, Zerilli:1970se}. Using this method, we can estimate the deviation from the Schwarzschild case
with high accuracy even for lower multipole number $\ell$.
The parametrized black hole QNM ringdown formalism can be applied to many models, especially around the static black holes in modified gravities~\cite{Cardoso:2019mqo, Cardoso:2018ptl, McManus:2019ulj, Tattersall:2019nmh, deRham:2020ejn}.

In general relativity, the gravitational perturbation around the Kerr black hole is described by the Teukolsky equation~\cite{Teukolsky:1973ha}.
Unfortunately, the parametrized QNM ringdown formalism apparently cannot to be applied to it\footnote{
Note that the Klein-Gordon equation around the slowly rotating Kerr black hole 
has been discussed in~\cite{Cardoso:2019mqo}.}
because the Teukolsky equation with the vanishing Kerr parameter $a=0$
does not reduce to the form of the Regge-Wheeler or Zerilli equation~\cite{Regge:1957td, Zerilli:1970se}.

In this paper, to avoid this problem, we rather focus on other formulations by Chandrasekhar and Detweiler \cite{ChandrasekharDetweiler:1976, Detweiler:1977} and by Sasaki and Nakamura~\cite{Sasaki:1981kj, Sasaki:1981sx, Nakamura:1981kk}.
These are related to the Teukolsky equation by highly non-trivial transformations of the master variables.
The key point is that 
both the Chandrasekhar-Detweiler equation and the Sasaki-Nakamura equation reduce to the Regge-Wheeler equation
in the non-rotating limit,\footnote{The Chandrasekhar-Detweiler equation also reduces to the Zerilli equation by flipping signs of a few parameters.}
and the parametrized black hole QNM ringdown formalism can be easily applied to them.
Using this property, we derive semi-analytic corrections\footnote{
Our approach is mainly based on analytical perturbative calculation 
around the Schwarzschild case. However, when we obtain a numerical value
of the QNM frequency, we need to use numerical results in 
the parametrized black hole QNM ringdown formalism~\cite{Cardoso:2019mqo, McManus:2019ulj}.
In this sence, we call our approach ``semi-analytic'' approach.} 
to the quasinormal modes in the slowly rotating limit up to the quadratic order of the Kerr parameter.
It turns out that both equations lead to perfectly the same result as expected.

Our approach is based on the parametrized black hole QNM ringdown formalism.
Though the QNMs of Kerr black holes have been well-studied especially by numerics {\it e.g.,} in~\cite{Berti:2009kk, Berti:2005ys},
there are still some motivations to further study them analytically:
(i) To obtain some insights from analytical expressions, {\it e.g.}, the dependence of 
the Kerr parameter $a$ and the azimuthal number $m$.
(ii) To provide a simple and quick approximate formula in the slow rotation regime, which is sufficiently accurate compared to fitting functions
derived from numerical results~\cite{Berti:2009kk, Berti:2005ys}.
(iii) To extend the discussion to rotating black holes in modified gravities in the future.
For this purpose, developing a semi-analytical derivation of the Kerr QNM with high accuracy for slow rotation case is needed.

The organization of this paper is as follows.
In Secs.II, III, we briefly review the parametrized black hole QNM ringdown formalism
and the Chandrasekhar-Detweiler equation. 
We derive the semi-analytic expression of quasinormal modes for slowly rotating black holes in Sec.IV, 
and compare our result with the previous numerical results~\cite{Berti:2009kk, Berti:2005ys} in Sec.V.
We summarize our results in Sec.VI.
In App.A, we discuss boundary conditions for QNMs.
In App.B, we show explicit results on the Sasaki-Nakamura equation.

\section{The parametrized black hole quasinormal mode ringdown formalism}
Let us review the parametrized black hole quasinormal mode (QNM) ringdown formalism~\cite{Cardoso:2019mqo, McManus:2019ulj, Kimura:2020mrh}.
In this formalism, we consider master equations with the following form:
\begin{align}
f \frac{d}{dr}\left(f \frac{d}{dr}\Phi\right)
+
(\omega^2 - f (V^{(0)} + \delta V))\Phi = 0,
\label{mastereq1}
\end{align}
with 
\begin{align}
f = 1 - \frac{r_H}{r},
\end{align}
where the constant $r_H$ denotes the location of the black hole horizon.
The background potential $V^{(0)}$ denotes the 
effective potential for the Zerilli or Regge-Wheeler equation~\cite{Regge:1957td, Zerilli:1970se}, {\it i.e.,}
\begin{align}
V^{(0)} &= V_+  = \frac{9 \lambda r_H^2 r + 3 \lambda^2 r_H r^2 + \lambda^2 (\lambda +2)r^3 + 9 r_H^3}{r^3(\lambda r + 3 r_H)^2},
\label{evenpotential}
\end{align}
with $\lambda = \ell^2 + \ell -2$ for the even parity perturbation, and
\begin{align}
V^{(0)} &= V_-  = \frac{\ell (\ell + 1)}{r^2} - \frac{3 r_H}{r^3},
\label{oddpotential}
\end{align}
for the odd parity perturbation.
We assume that $\delta V$ takes the form
\begin{align}
\delta V = \delta V_\pm &= 
\frac{1}{r_H^2}\sum_{j = 0}^\infty \alpha_j^\pm \left(\frac{r_H}{r}\right)^j,
\label{deltav}
\end{align}
where $\alpha_j^\pm$ are small parameters.
At the quadratic order of the small parameters $\alpha_j^\pm$, the QNM frequency of Eq.~\eqref{mastereq1}
is given by
\begin{align}
\omega_{\rm QNM} &= \omega_0  + \sum_{j = 0}^\infty \alpha_j^\pm e_j^\pm
+
 \sum_{j,k = 0}^\infty \alpha_j^\pm \alpha_k^\pm e_{jk}^\pm 
+
{\cal O}(\alpha^3),
\label{qnmomegaformula1}
\end{align}
where $\omega_0 = 2\Omega_0/r_H$ and
$\Omega_0$ denotes the dimensionless QNM frequency for the Schwarzschild black hole, {\it e.g.,}
\begin{align}
\Omega_0 = 0.3736716844180418...- i0.0889623156889357... ~~ ({\rm for} ~\ell = 2~{\rm fundamental ~ mode).}
\end{align}
We should note that the coefficients 
$e_j^\pm$ and $e_{jk}^\pm$ do not depend on the small parameters $\alpha_j^\pm$,
and thus our task is to know them.
Their numerical data are found in~\cite{Cardoso:2019mqo, McManus:2019ulj, HatsudaKimurainprep}.\footnote{
In~\cite{HatsudaKimurainprep}, we recalculated the coefficients $e_j^\pm$ and $e_{jk}^\pm$.
We used the recursion relations for $e_j^-$~\cite{Kimura:2020mrh} and their higher order extensions 
to evaluate the numerical error. We confirmed that the errors are ${\cal O}(10^{-15})$.}

\section{The Chandrasekhar-Detweiler equation}
The Chandrasekhar-Detweiler equation~\cite{ChandrasekharDetweiler:1976, Detweiler:1977} was obtained by a transformation of the Teukolsky equation.
It is given by\footnote{
The relation between $X$ and the Teukolsky variable $R$ can be seen in~\cite{ChandrasekharDetweiler:1976, Detweiler:1977, Nakamura:2016gri}.
}
\begin{align}
\left(\frac{d^2}{dr_*^2} + (\omega^2 -V_{\rm CD}) \right)X = 0,
\label{eq:cdeq}
\end{align}
where $r_*$ is defined by\footnote{
Note that $r_*$ can be written explicitly as
\begin{align}
r_* = r + \frac{2 M r_+}{r_+ - r_-}\ln \frac{r-r_+}{2M}
-
\frac{2 M r_-}{r_+ - r_-} \ln \frac{r-r_-}{2 M}.
\end{align}
}
\begin{align}
\frac{dr_*}{dr} &= \frac{r^2 + a^2}{\Delta},
\label{eq:rstardef}
\\
\Delta &= r^2 - 2 M r + a^2.
\end{align}
The potential $V_{\rm CD}$ is
\begin{align}
V_{\rm CD} &= \omega^2 + {\cal V},
\\
{\cal V} &=
\frac{-K^2 + \Delta \lambda}{(r^2 + a^2)^2}
+
\frac{2\Delta (r^3 M + a^4)}{r^2(r^2 + a^2)^3}
+
\frac{3 a^2 \Delta^2}{(r^2 + a^2)^4}
\notag\\&\quad-
\frac{4 \lambda \rho^2 \Delta 
(
-2 \lambda \rho^2 (r^2 - a^2) + 2 r (r M - a^2)(4\lambda r + 6 M + \kappa_2)
)
}{r^2(r^2 + a^2)^2 (2 \lambda r^2 + (6 M + \kappa_2)r - 2\lambda(a^2 + am/\sigma))^2},
\end{align}
where $K, \rho, \kappa_2$ are given by
\begin{align}
K &= (r^2 + a^2) \omega - am,
\\
\rho^2 &= r^2 + a^2 + \frac{a m}{\sigma},
\\
\kappa_2 &= \pm \sqrt{36 M^2 - 2 \lambda\left(\left(a^2 + \frac{am}{\sigma}\right)(5\lambda + 6) - 12 a^2\right) 
+
2 b_2 \lambda(\lambda + 2)},
\end{align}
with
\begin{align}
\sigma =-\omega, \qquad b_2 &= \pm 3 \left(a^2 + \frac{am}{\sigma}\right).
\end{align}
The parameters $M, a, m$ denote the mass parameter, the Kerr parameter and the azimuthal number, respectively.
Hereafter, we choose the minus signs of $b_2$ and $\kappa_2$.\footnote{
The minus signs lead to the Regge-Wheeler potential in $a \to 0$, while the plus signs to the Zerilli potential.}
In the calculation of QNM frequencies for slowly rotating Kerr black holes,
we mainly use $r_\pm = M \pm \sqrt{M^2 - a^2}$ as 
fundamental variables rather than $M$ and $a$.
In the slow rotation limit $a \to 0$, we have $r_+ \sim 2M$, $r_- \sim a^2/(2M)$.
Note that $M$ and $a$ can be written by $r_\pm$ as
\begin{align}
M=\frac{r_++r_-}{2},\qquad a=\sqrt{r_+ r_-}.
\end{align}

We can easily check that the Chandrasekhar-Detweiler equation reduces to the Regge-Wheeler equation in the limit of $a \to 0$.
Thus it is expected to compute the QNMs for the slow rotation case (small $a$ or small $\sqrt{r_-}$)
by using the parametrized black hole QNM ringdown formalism.
To show it, we introduce a new master variable $\tilde{X}$
\begin{align}
\tilde{X} = \sqrt{Z} X,
\end{align}
with $Z = r(r-r_-)/(r^2 + a^2)$.
Then the Chandrasekhar-Detweiler equation~\eqref{eq:cdeq} becomes
\begin{align}
& f \frac{d}{dr}
\left(f \frac{d}{dr}\right) \tilde{X}
+\left( \frac{\omega^2}{Z^2}
-
f \tilde{V}
\right)\tilde{X} = 0,
\label{cdeqtildex}
\end{align}
with
\begin{align}
f &= 1- \frac{r_+}{r},
\\
\tilde{V} &= \frac{V_{\rm CD}}{f Z^2} - \frac{f (\partial_r Z)^2 - 2 Z \partial_r(f \partial_r Z)}{4 Z^2}.
\end{align}
As shown in the next section, taking the series expansion of $\sqrt{r_-}$ to the above equation,
we can rewrite the master equation as the form so that the parametrized black hole QNM ringdown formalism can be applied.
We note that the QNM boundary condition of $\tilde{X}$
is same as $X$ because the function $Z$ is regular at $r = r_+$ and $\infty$.
The asymptotic behavior of $\tilde{X}$ for slowly rotating case is discussed in Appendix.~\ref{qnmbc}.

In the above equations, $\lambda$ is an eigenvalue of the differential equation for the angular part
\begin{align}
\frac{1}{\sin\theta}\frac{d}{d\theta}\left(\sin\theta \frac{dS(\theta)}{d\theta}\right)
+
\left[-
a^2 \omega^2 \sin^2\theta
-
\frac{(m-2\cos\theta)^2}{\sin^2\theta}
+
4 a \omega \cos\theta -2 + 2ma\omega + \lambda
\right]S(\theta) = 0,
\label{lambdaequation}
\end{align}
with the regular boundary condition at $\theta = 0, \pi$.
Defining $x:= \cos\theta$, Eq.\eqref{lambdaequation} takes the form
\begin{align}
\frac{d}{dx}\left((1-x^2)\frac{dS}{dx}\right)
+
\left[
(a \omega x)^2
+
4 a \omega x 
-
2
+
(\lambda   + 2 m a \omega - a^2\omega^2)
-
\frac{(m-2 x)^2}{1-x^2}
\right]S = 0.
\label{eqangularpart}
\end{align}
In \cite{Berti:2005gp}, the eigenvalue $\lambda$ for the above equation 
was discussed in detail.\footnote{
Note that Eq.~\eqref{eqangularpart} corresponds to Eq.(2.1) in \cite{Berti:2005gp} with $s = -2, c = a \omega,   A = \lambda   + 2 m a \omega - a^2\omega^2$.}
For small $\sqrt{r_-/ r_+} ~(= a/r_+)$, the eigenvalue $\lambda$ behaves as
\begin{align}
\lambda &= \lambda_0 + \lambda_1 m r_+ \omega \sqrt{\frac{r_-}{r_+}}
+
\lambda_2 (r_+ \omega)^2 \frac{r_-}{r_+}
+ {\cal O }(r_-^{3/2}),
\label{eqlambda}
\end{align}
with
\begin{align}
\lambda_0 &= \ell^2 + \ell - 2,
\\
\lambda_1 &= -2 - \frac{8}{\ell(\ell + 1)},
\\
\lambda_2 &= h_{\ell + 1,m} - h_{\ell,m},
\\
h_{\ell,m} &=  \frac{(\ell^2 -m^2)(\ell^2 - 4)^2}{2(\ell^2 - 1/4)\ell^3}.
\end{align}
Note that when we calculate QNM frequencies for slowly rotating Kerr black holes, 
$\omega$ in Eq.~\eqref{eqlambda} also can be expanded as 
\begin{align}
\omega = \omega_0 + \omega_1 \sqrt{\frac{r_-}{r_+}} + \omega_2 \frac{r_-}{r_+} + {\cal O }(r_-^{3/2}).
\end{align}
We determine the coefficients $\omega_1$ and $\omega_2$ semi-analytically in the next section.

\section{QNM frequencies for slowly rotating Kerr black holes}
Now we show that the Chandrasekhar-Detweiler equation allows us to apply the parametrized black hole QNM ringdown formalism to slowly rotating black holes.

\subsection{First order expression}
At the first order of $\sqrt{r_-/r_+}$, we can rewrite the Chandrasekhar-Detweiler equation~\eqref{cdeqtildex} in the form
\begin{align}
& f \frac{d}{dr}
\left(f \frac{d}{dr}\right) \tilde{X}
+\left(\left(\omega - \frac{m}{r_+}\sqrt{\frac{r_-}{r_+}}\right)^2 
- f(V_{-} + \delta V 
)\right)\tilde{X} = 0,
\label{mastereq1storder}
\end{align}
with
\begin{align}
V_- &= \frac{\ell(\ell +1)}{r^2} - \frac{3r_+}{r^3},
\label{vminus}
\\
\delta V &= \frac{1}{r_+^2}\sum_{j = 0}^5 \alpha_j^- \left(\frac{r_+}{r}\right)^j,
\end{align}
where $\alpha_j^- = \alpha_j^{({\rm 1st})}  \sqrt{r_-/r_+}$ and the explicit forms of $\alpha_j^{({\rm 1st})}$ are given by 
\begin{align}
\alpha_0^{({\rm 1st})} &= -2 m \omega_0 r_+ ,
\label{eqalpha0}
\\
\alpha_1^{({\rm 1st})} &= -2 m \omega_0 r_+ ,
\\
\alpha_2^{({\rm 1st})} &=  
 m \omega_0 r_+ \lambda_1,
\\
\alpha_3^{({\rm 1st})} &= 
 \frac{8m(3 + 2\lambda_0)}{3 \omega_0 r_+},
\\
\alpha_4^{({\rm 1st})} &= 
- \frac{4 m (5+2\lambda_0)}{\omega_0 r_+},
\\
\alpha_5^{({\rm 1st})} &= \frac{12 m}{\omega_0 r_+}.
\label{eqalpha5}
\end{align}
We note that the master equation~\eqref{mastereq1storder} is equivalent to that in \cite{Pani:2013pma} up to an ambiguity of the effective potential at the first order of $a$~\cite{Kimura:2020mrh}.
Using the parametrized black hole QNM ringdown formalism, we immediately obtain the semi-analytic QNM frequency of Kerr black hole at the first order of $\sqrt{r_-/r_+}$
\begin{align}
\omega_{\rm QNM}^{\rm Kerr}  = \frac{2\Omega_0}{r_+} + \biggl( \frac{m}{r_+} +\sum_{j = 0}^5 \alpha_j^{({\rm 1st})} e_j^- \biggr) \sqrt{\frac{r_-}{r_+}} + {\cal O}(r_-).
\label{qnm1storder}
\end{align}
For the fundamental mode with $\ell = 2$, the QNM frequency Eq.~\eqref{eq:qnmKerr2ndorder}
at ${\cal O}(a)$ becomes
\begin{align}
M \omega_{\rm [1st]}^{\rm QNM} &=
(0.3736716844180418 - 
   0.0889623156889357 i) 
\notag\\ & + (0.0628830795083 + 0.0009979348536  i) \frac{ma}{M}
+
{\cal O}(a^2),
\label{eq:kerrqnm1stformula}
\end{align}
where we used $r_\pm = M \pm \sqrt{M^2 - a^2}$ and the numerical data of $e_{j}^-$~\cite{Cardoso:2019mqo, Hatsuda:2019eoj, HatsudaKimurainprep}.
We note that we can obtain the same result from the Sasaki-Nakamura equation if we use the recursion relation for $e_j^-$~\cite{Kimura:2020mrh} (see Appendix~\ref{sec:sasakinakamura}).

\subsection{Second order expression}
The extension to the quadratic order is straightforward. At the second order of $\sqrt{r_-/r_+}$, the Chandrasekhar-Detweiler equation~\eqref{cdeqtildex} becomes
\begin{align}
& f \frac{d}{dr}
\left(f \frac{d}{dr}\right) \tilde{X}
+\left(\left(\omega - \frac{m}{r_+}\sqrt{\frac{r_-}{r_+}} 
+
 2 \omega_0 \frac{r_-}{r_+}
\right)^2 
- f(V_{-} + \delta V)\right)\tilde{X} = 0,
\end{align}
with
\begin{align}
\delta V &= \frac{1}{r_+^2}\sum_{j = 0}^7 \alpha_j^- \left(\frac{r_+}{r}\right)^j,
\end{align}
where
$\alpha_j^- = \alpha_j^{({\rm 1st})} \sqrt{r_-/r_+} + \alpha_j^{({\rm 2nd})} r_-/r_+$.
The first order corrections $\alpha_j^{({\rm 1st})}$ 
are same as Eqs.~\eqref{eqalpha0}-\eqref{eqalpha5} (but $\alpha_6^{({\rm 1st})} = \alpha_7^{({\rm 1st})} = 0$),
and the second order corrections $\alpha_j^{({\rm 2nd})}$ are given by 
\begin{align}
\alpha_0^{({\rm 2nd})} &= m^2-2 m r_+ \omega_1+4 r_+^2 \omega_0^2,
\\
\alpha_1^{({\rm 2nd})} &= m^2-2 m r_+ \omega_1+2 r_+^2 \omega_0^2,
\\
\alpha_2^{({\rm 2nd})} &= m^2+\lambda_1 m r_+ \omega_1+\lambda_2 r_+^2 \omega_0^2,
\\
\alpha_3^{({\rm 2nd})} &= -\frac{m^2 (16 \lambda_0 (2 \lambda_0+3)^2-9 (16 \lambda_1+3) r_+^2 \omega_0^2)+72 (2 \lambda_0+3) m r_+ \omega_1+9 (13 \lambda_0+6) r_+^2 \omega_0^2}{27 r_+^2 \omega_0^2},
\\
\alpha_4^{({\rm 2nd})} &=\frac{16 m^2 ((\lambda_0+3) (2 \lambda_0+3)^2-9 \lambda_1 r_+^2 \omega_0^2)+72 (2 \lambda_0+5) m r_+ \omega_1+9 (16 \lambda_0+25) r_+^2 \omega_0^2}{18 r_+^2 \omega_0^2},
\\
\alpha_5^{({\rm 2nd})} &=-\frac{4 (2 (2 \lambda_0+3) (8 \lambda_0+27) m^2+27 m r_+ \omega_1+27 \
r_+^2 \omega_0^2)}{9 r_+^2 \omega_0^2},
\\
\alpha_6^{({\rm 2nd})} &= \frac{4 (26 \lambda_0+51) m^2}{3 r_+^2 \omega_0^2},
\\
\alpha_7^{({\rm 2nd})} &= -\frac{20 m^2}{r_+^2 \omega_0^2},
\end{align}
where $\omega_1$ is the first order correction derived in the previous subsection:
\begin{align}
\omega_1  = \frac{m}{r_+}+ \sum_{j = 0}^5 \alpha_j^{({\rm 1st})} e_j^-.
\end{align}
Thus, the semi-analytic QNM frequency at the second order of $\sqrt{r_-/r_+}$ finally becomes
\begin{align}
\omega_{\rm QNM}^{\rm Kerr}  &= \frac{2\Omega_0}{r_+}
+ \frac{m}{r_+}\sqrt{\frac{r_-}{r_+}}
 - 2\omega_0 \frac{r_-}{r_+}
+ \sum_{j = 0}^7 \alpha_j^{-} e_j^- 
+ \sum_{j,k = 0}^5 \alpha_j^{-} \alpha_k^{-} e_{jk}^- 
+ {\cal O}(r_-^{3/2}), \\
&=\frac{2\Omega_0}{r_+}+ \biggl( \frac{m}{r_+} +\sum_{j = 0}^5 \alpha_j^{({\rm 1st})} e_j^- \biggr) \sqrt{\frac{r_-}{r_+}} 
\notag \\
&\quad+\biggl( -2\omega_0+\sum_{j=0}^7 \alpha_j^{(\rm 2nd)} e_j^- +\sum_{j,k=0}^5 \alpha_j^{({\rm 1st})}\alpha_k^{({\rm 1st})} e_{jk}^- \biggr)\frac{r_-}{r_+}
+{\cal O}(r_-^{3/2}).
\label{eq:qnmKerr2ndorder}
\end{align}
We should note that, for $m = 0$,
the first order correction terms vanish (see Eqs.~\eqref{eqalpha0}-\eqref{eqalpha5}), but 
the second order correction does not.
For the fundamental mode with $\ell = 2$, the QNM frequency Eq.~\eqref{eq:qnmKerr2ndorder}
at ${\cal O}(a^2)$ becomes
\begin{align}M \omega_{\rm [2nd]}^{\rm QNM} &=
(0.3736716844180418 - 
   0.0889623156889357 i) 
\notag\\ & + (0.0628830795083 + 0.0009979348536  i) \frac{ma}{M}
\notag\\ &+ 
 \Big((0.03591312868219 + 0.00638178925048  i) 
\notag\\ &
+ (0.00895679029 - 0.00029122129  i) m^2\Big) \left(\frac{a}{M}\right)^2
+
{\cal O}(a^3),
\label{eq:kerrqnm2ndformula}
\end{align}
where we used $r_\pm = M \pm \sqrt{M^2 - a^2}$ and the numerical data of $e_{j}^-$ and $e_{jk}^-$~\cite{Cardoso:2019mqo, McManus:2019ulj, HatsudaKimurainprep}.
We confirmed that the Sasaki-Nakamura equation reproduces the same result in the end.

\section{Comparison with numerical calculation}

\begin{figure}
\begin{center}
\includegraphics[width=3.8cm]{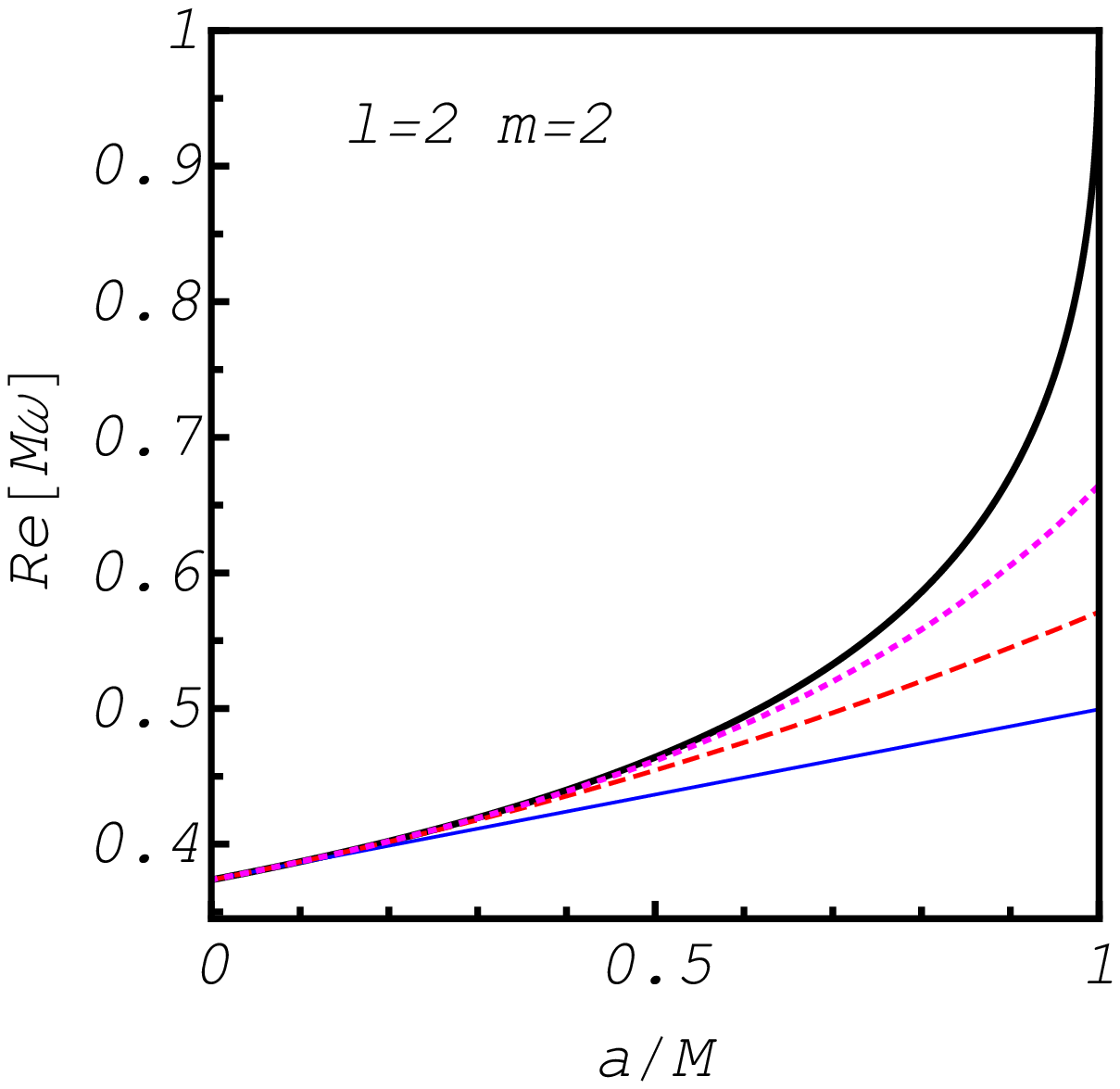}~~
\includegraphics[width=4.1cm]{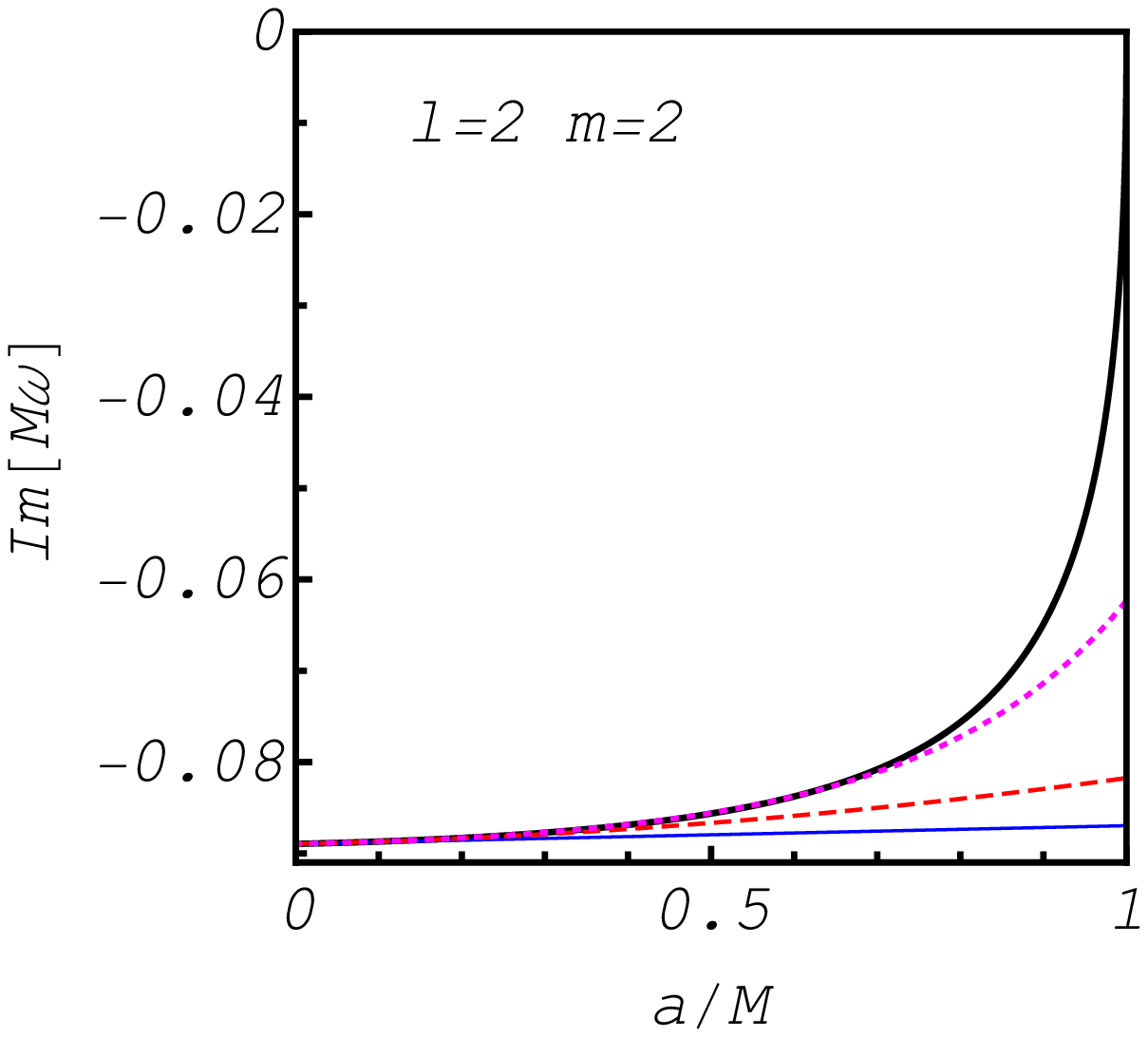}~~
\includegraphics[width=3.8cm]{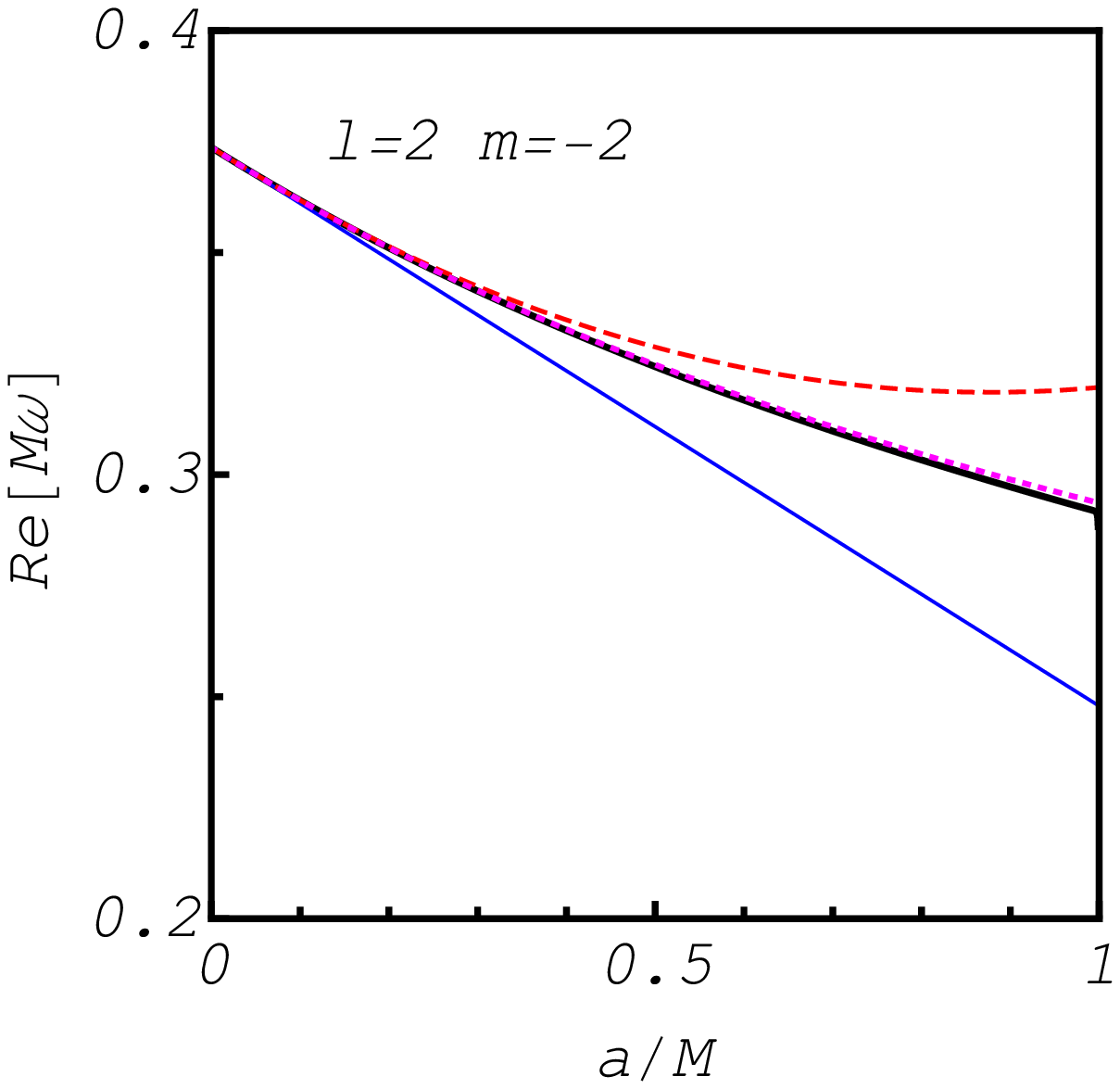}~~
\includegraphics[width=4.3cm]{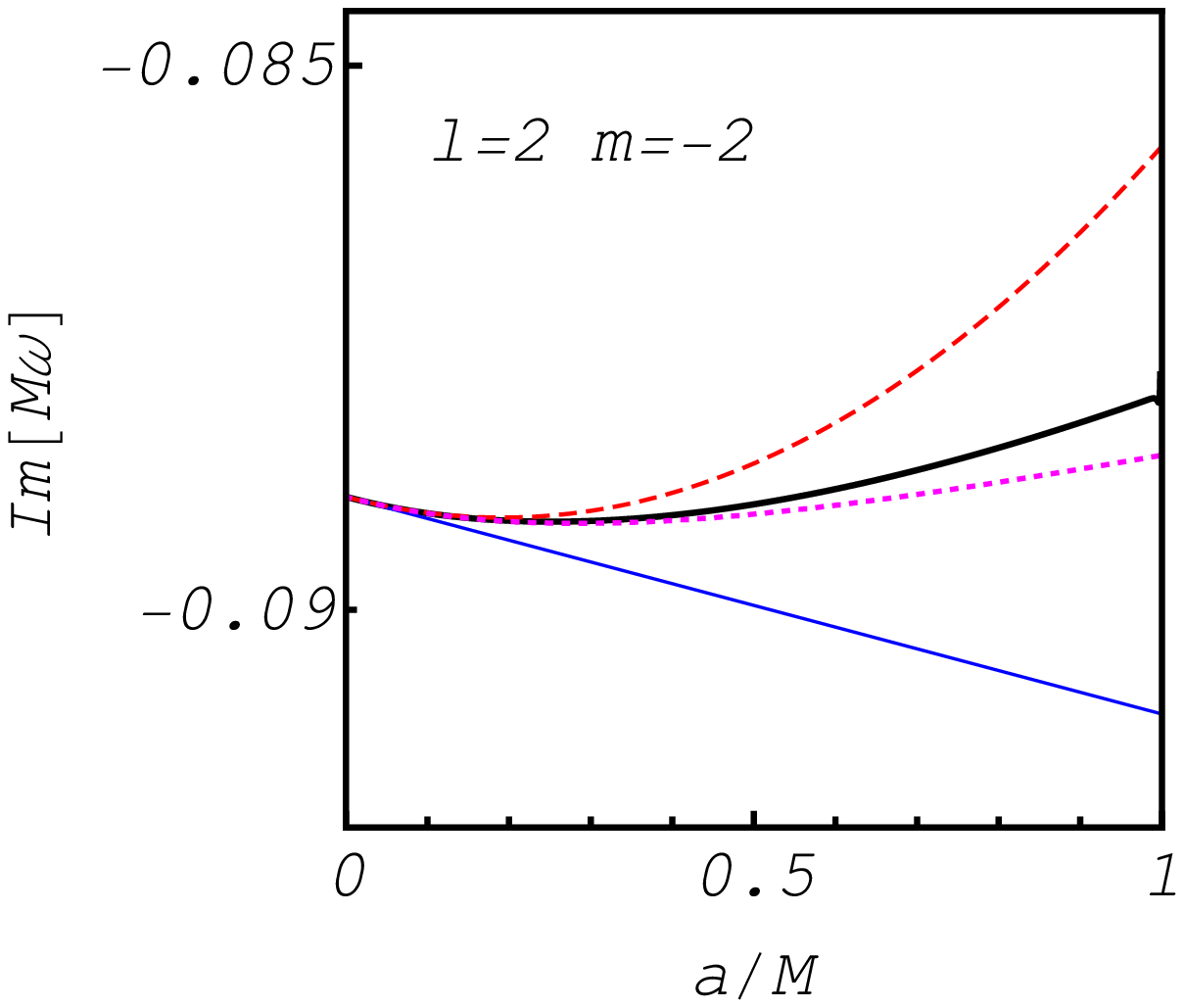}
\\
\includegraphics[width=3.8cm]{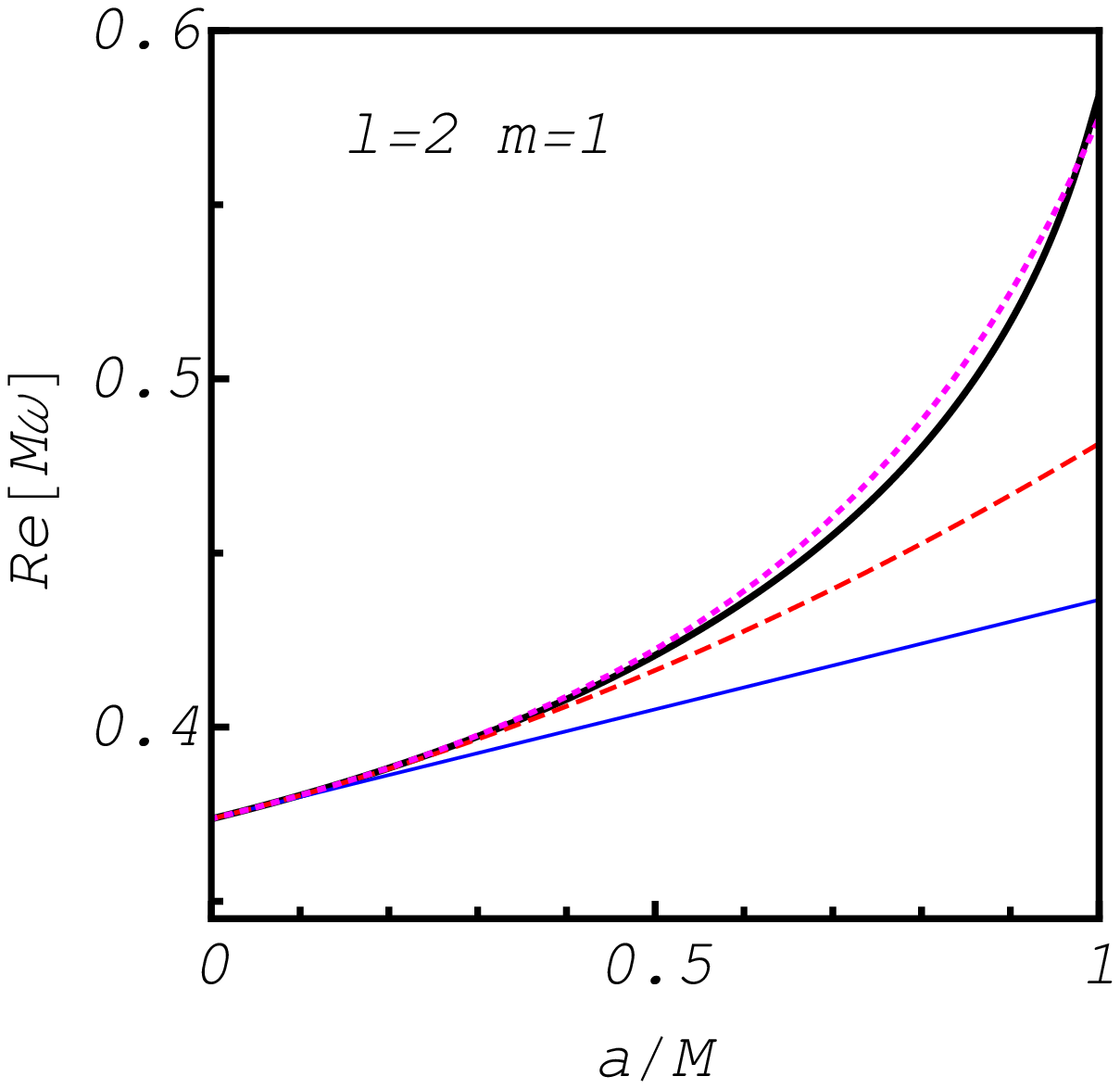}~~
\includegraphics[width=4.2cm]{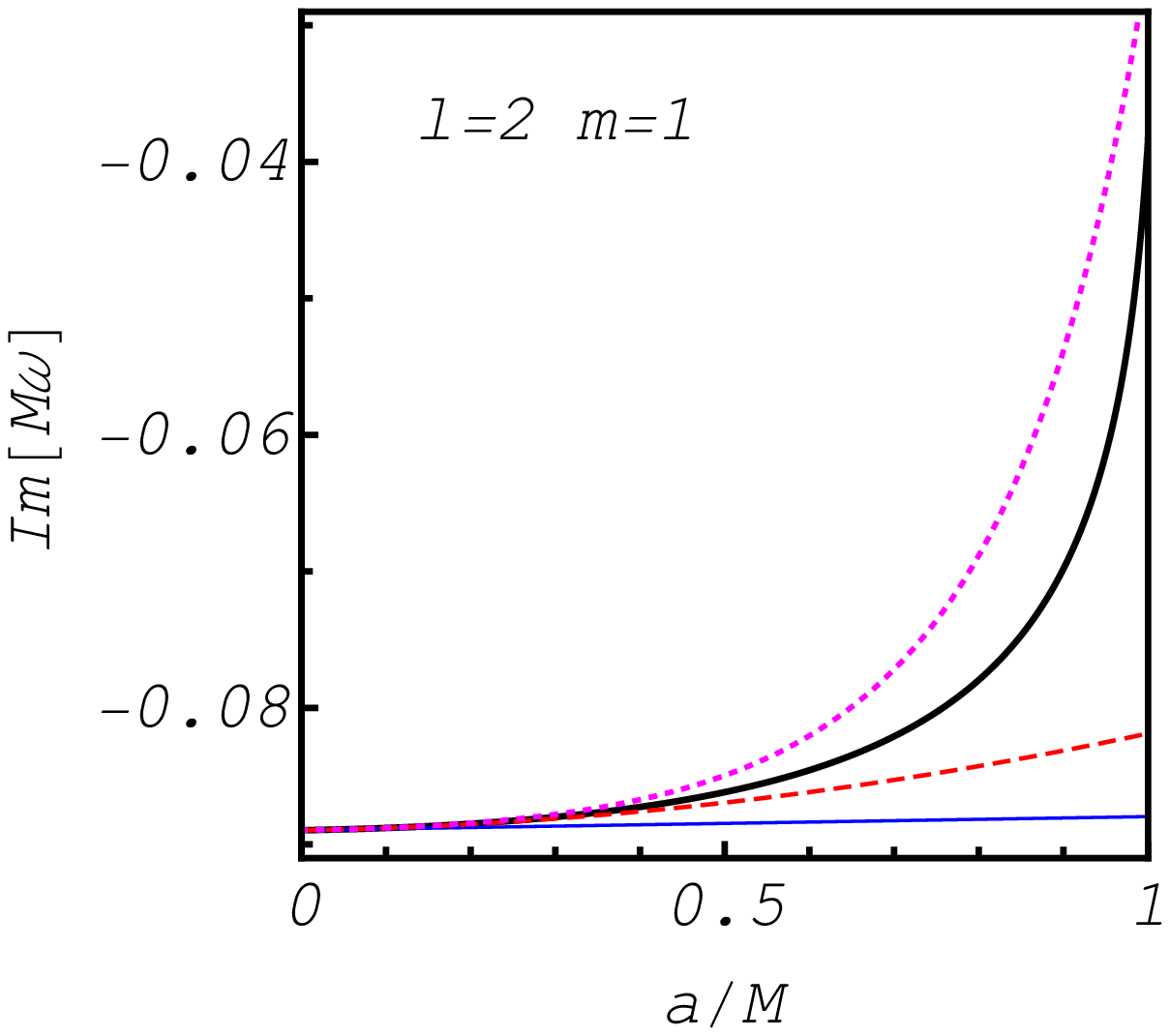}~~
\includegraphics[width=3.8cm]{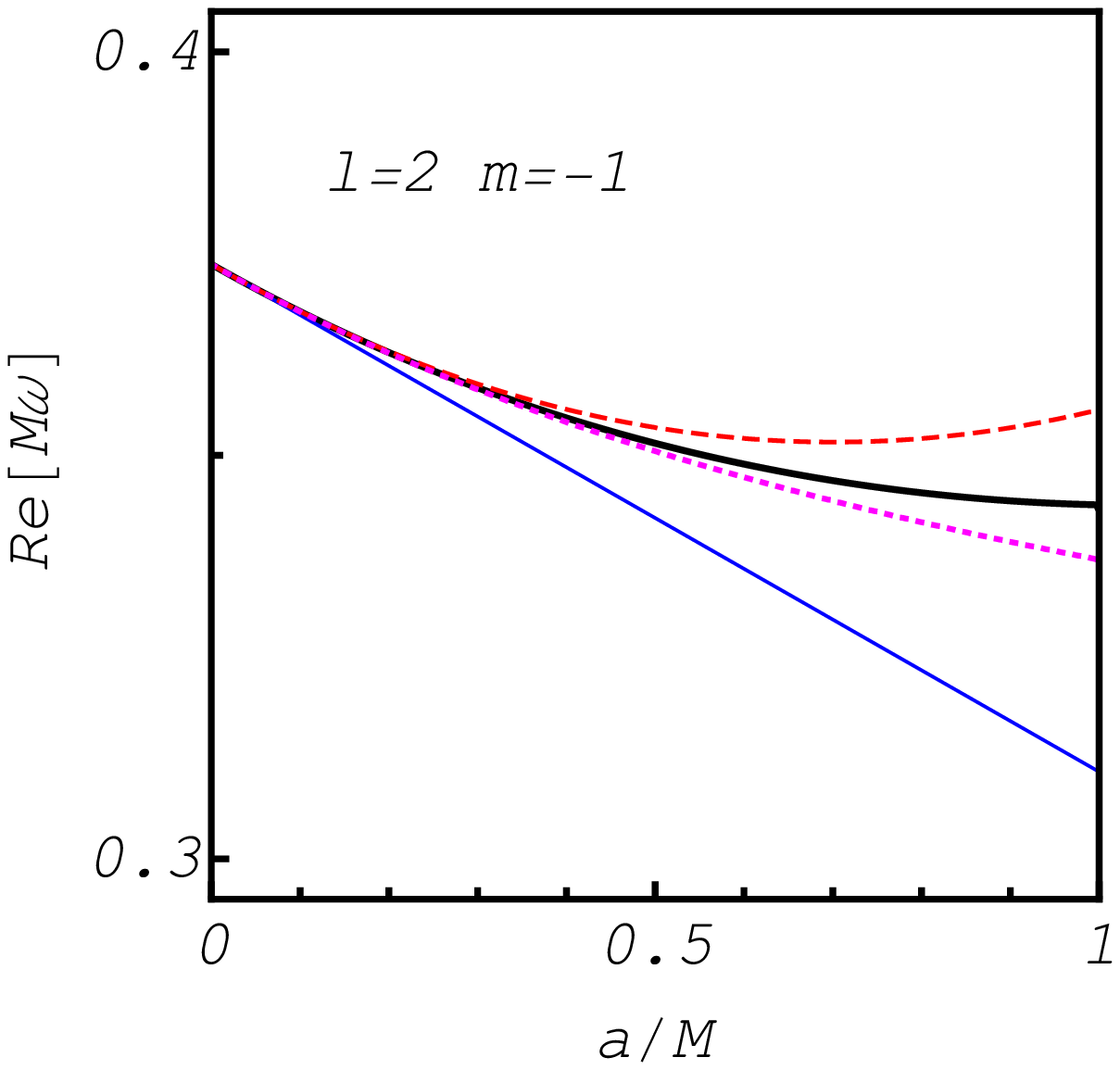}~~
\includegraphics[width=4.2cm]{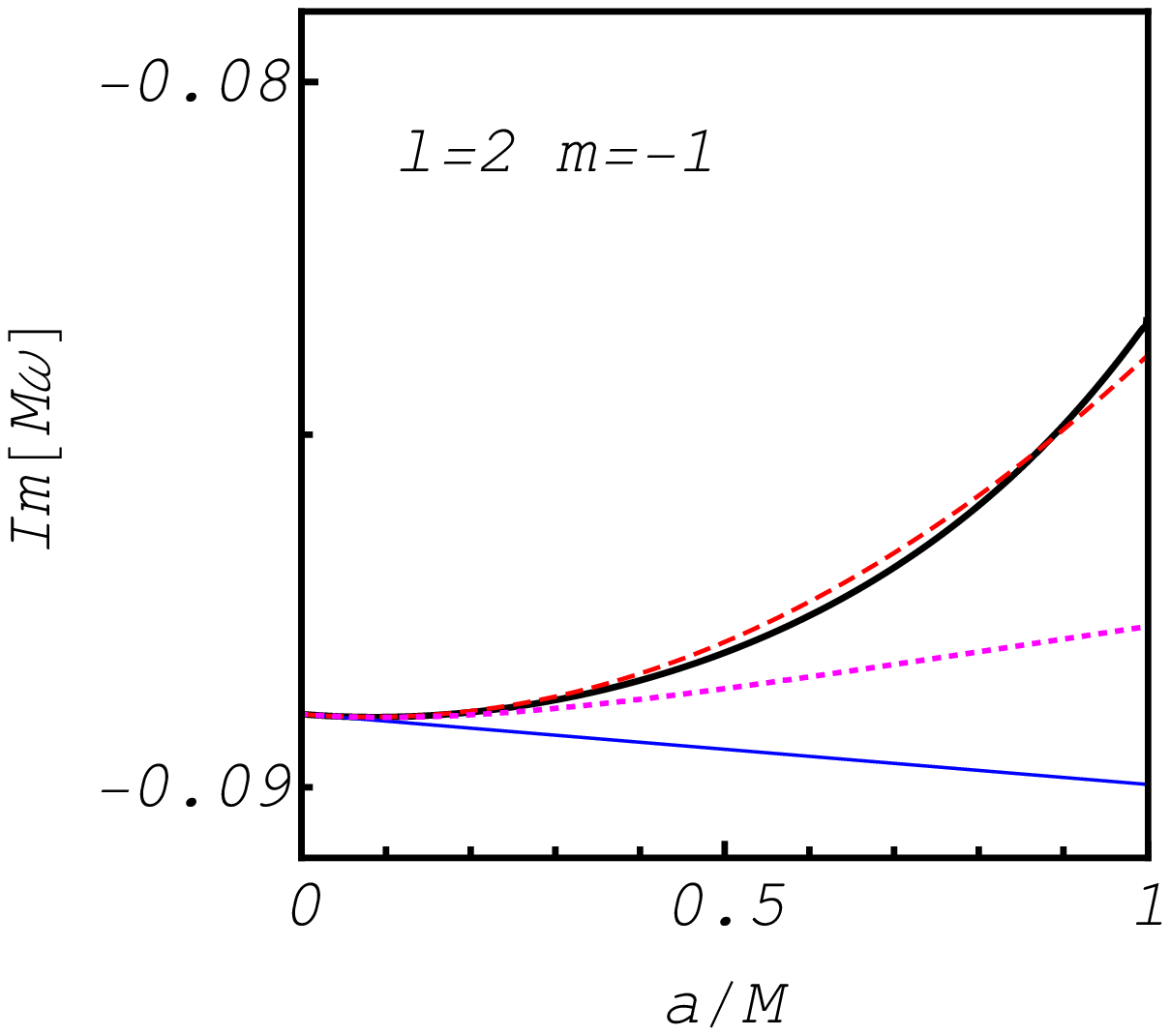}
\\
\includegraphics[width=3.9cm]{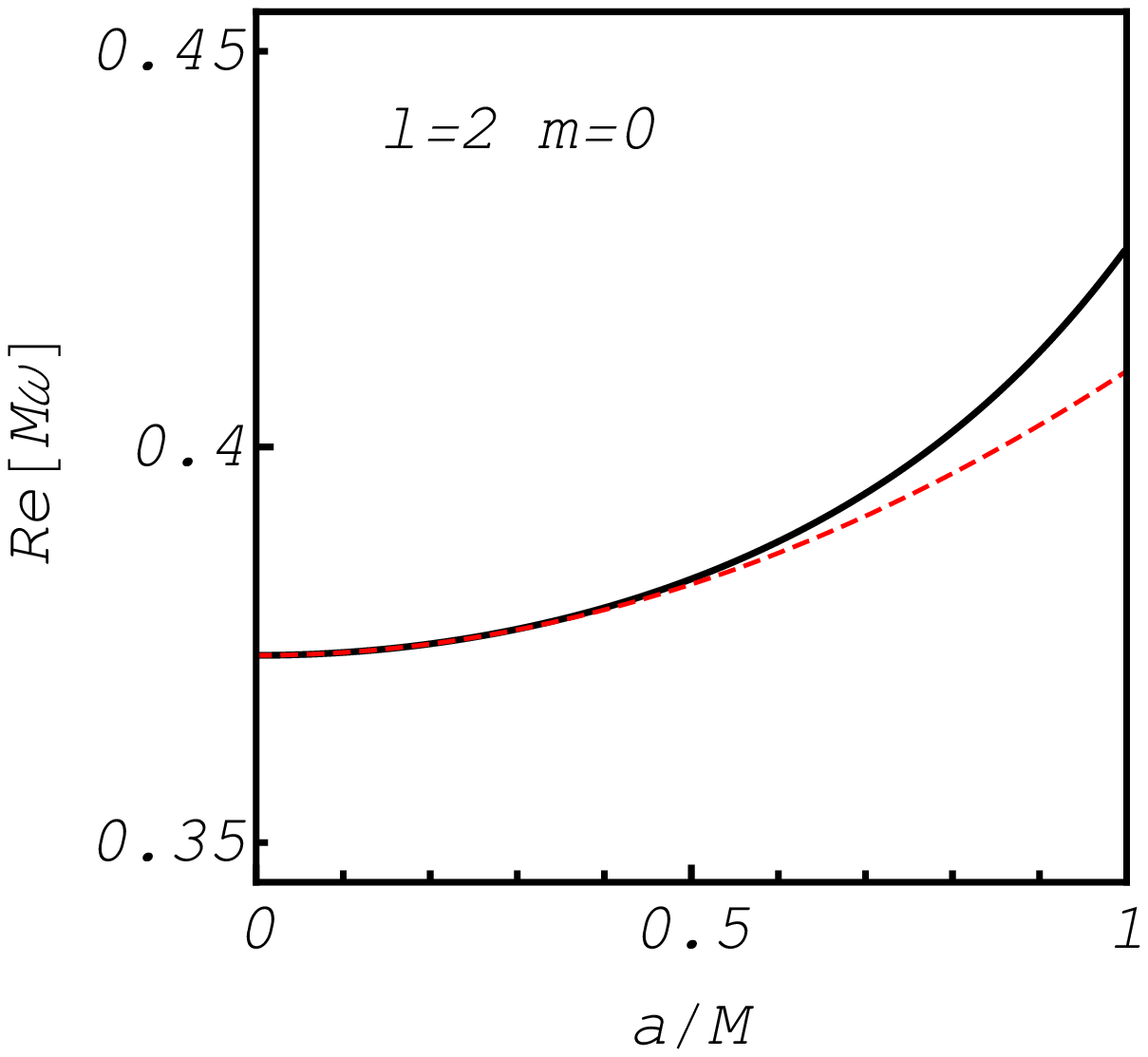}~~
\includegraphics[width=4.2cm]{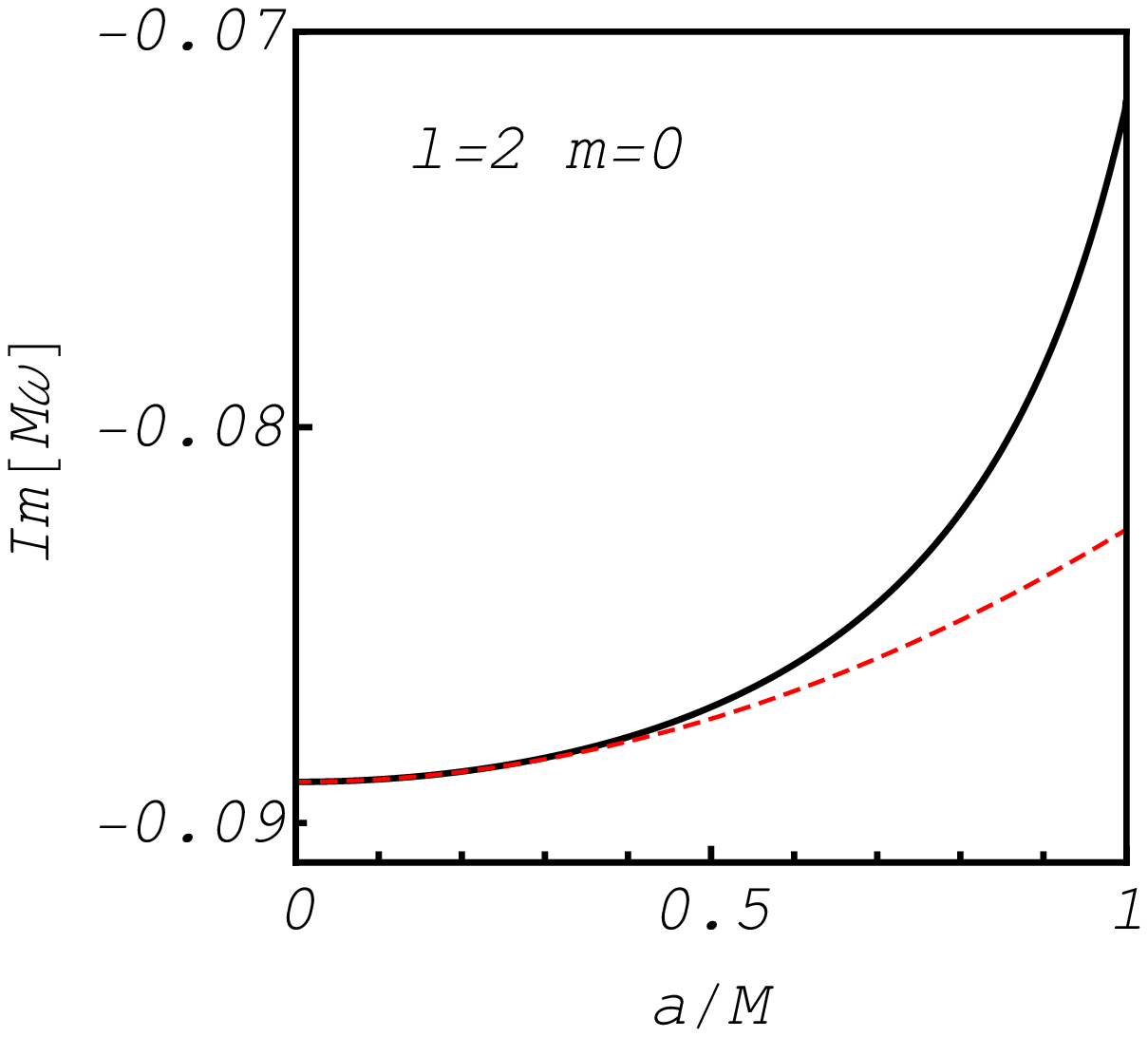}
\end{center}
\caption{\label{fig1}
The QNM frequencies of Kerr black holes for $\ell = 2$ with $m = 0, \pm 1, \pm 2$.
Thick (black), solid (blue), dashed (red), dotted (magenda) curves denote
the numerical calculation in~\cite{Berti:2005ys, Berti:2009kk, ringdowndata}, the first order expression, the second order expression and the $[1/1]$ Pad\'e approximant, respectively.
}
\end{figure}

In this section, we 
compare our semi-analytic expression of quasinormal modes for slowly rotating Kerr black holes, 
Eqs.~\eqref{eq:kerrqnm1stformula} and~\eqref{eq:kerrqnm2ndformula},
with numerical results in~\cite{Berti:2005ys, Berti:2009kk}.
In addition to Eqs.~\eqref{eq:kerrqnm1stformula} and~\eqref{eq:kerrqnm2ndformula}, we also use the Pad\'e approximant\footnote{
For $F(x) = A + B x + C x^2$ with $B \neq 0$, its Pad\'e approximant of degree $[1/1]$
is explicitly given by
\begin{align}
F^{[1/1]}(x):=\frac{A + (B^2 - AC)x/B}{1 - Cx/B}.
\end{align}
This is a rational function whose Taylor series is the same as the original function up to the second order.
}
from the second order expression Eq.~\eqref{eq:kerrqnm2ndformula}.
In Fig.~\ref{fig1}, we plot 
the QNM frequencies of Kerr black holes for $\ell = 2$ fundamental modes with $m = 0, \pm 1, \pm 2$.

\begin{figure}
\begin{center}
\includegraphics[width=4cm]{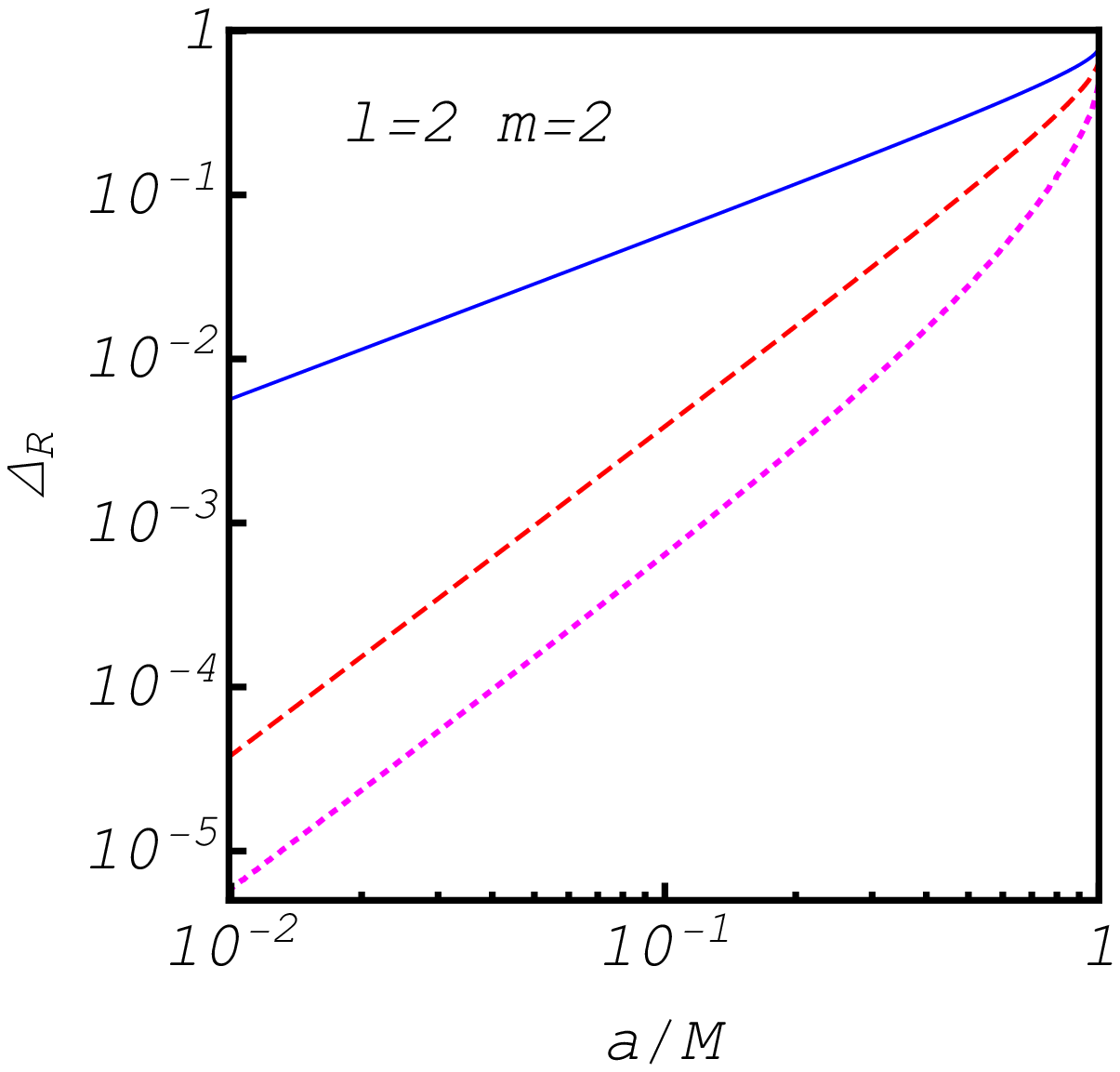}~~
\includegraphics[width=4cm]{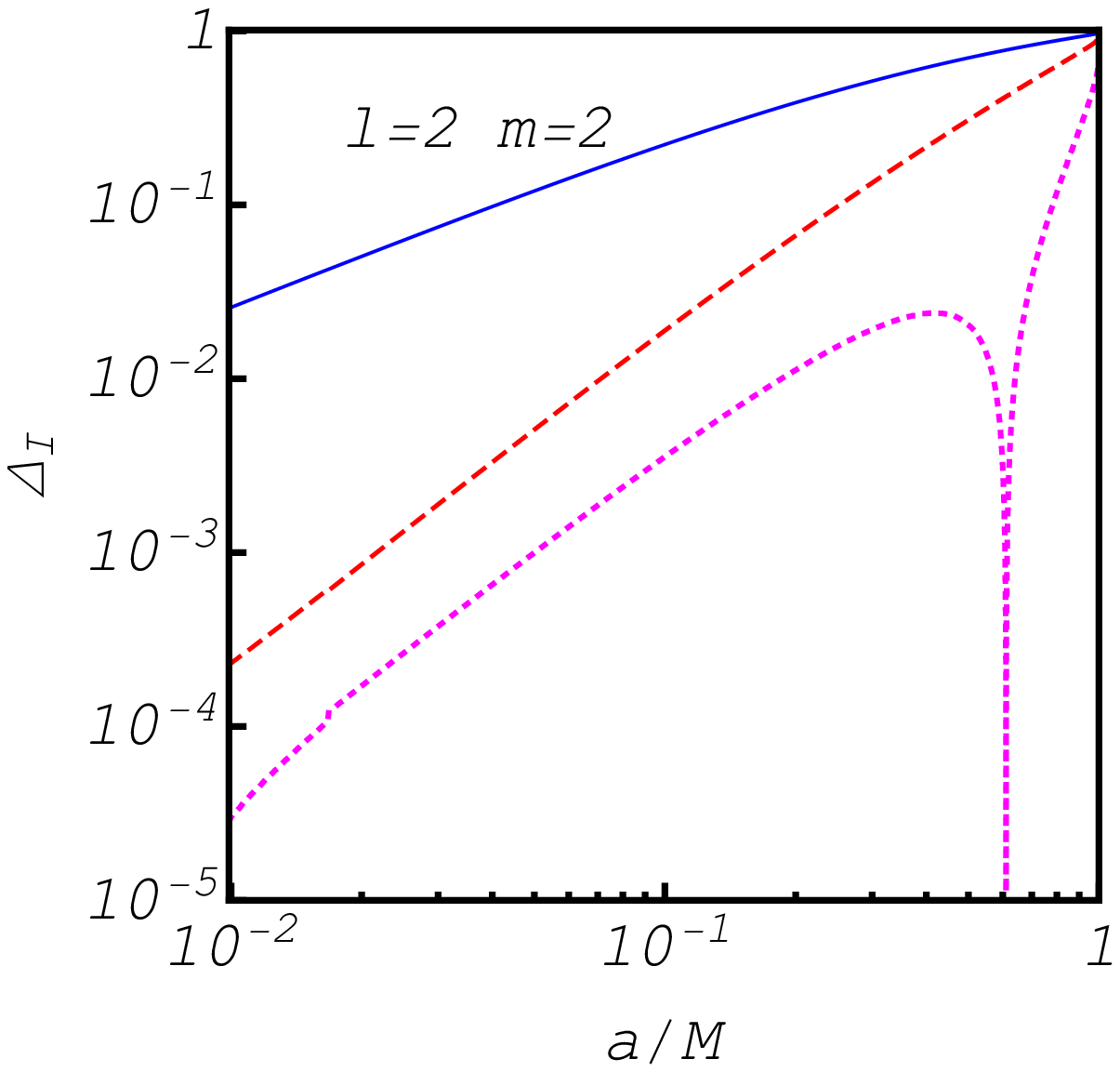}~~
\includegraphics[width=4cm]{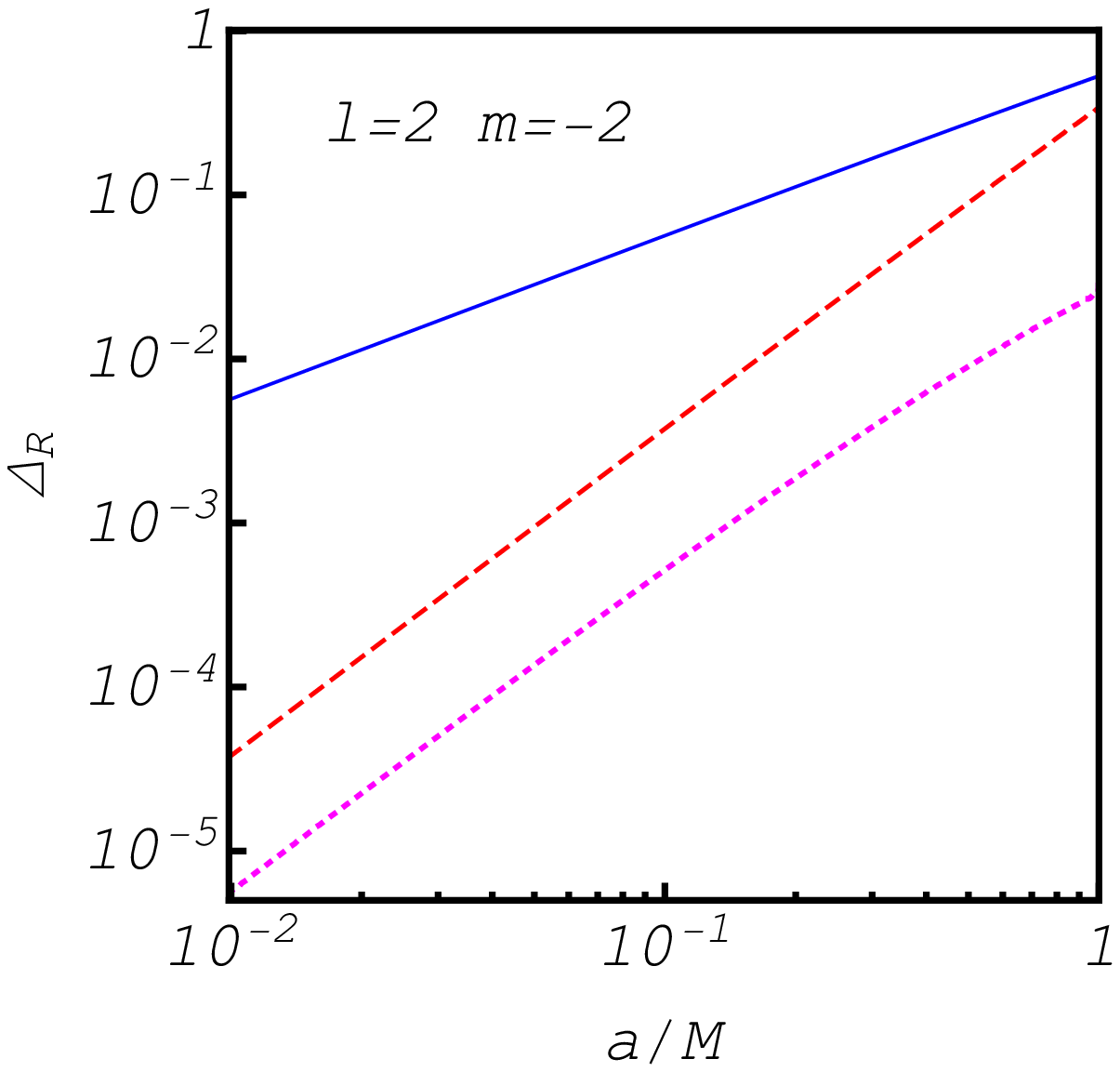}~~
\includegraphics[width=4cm]{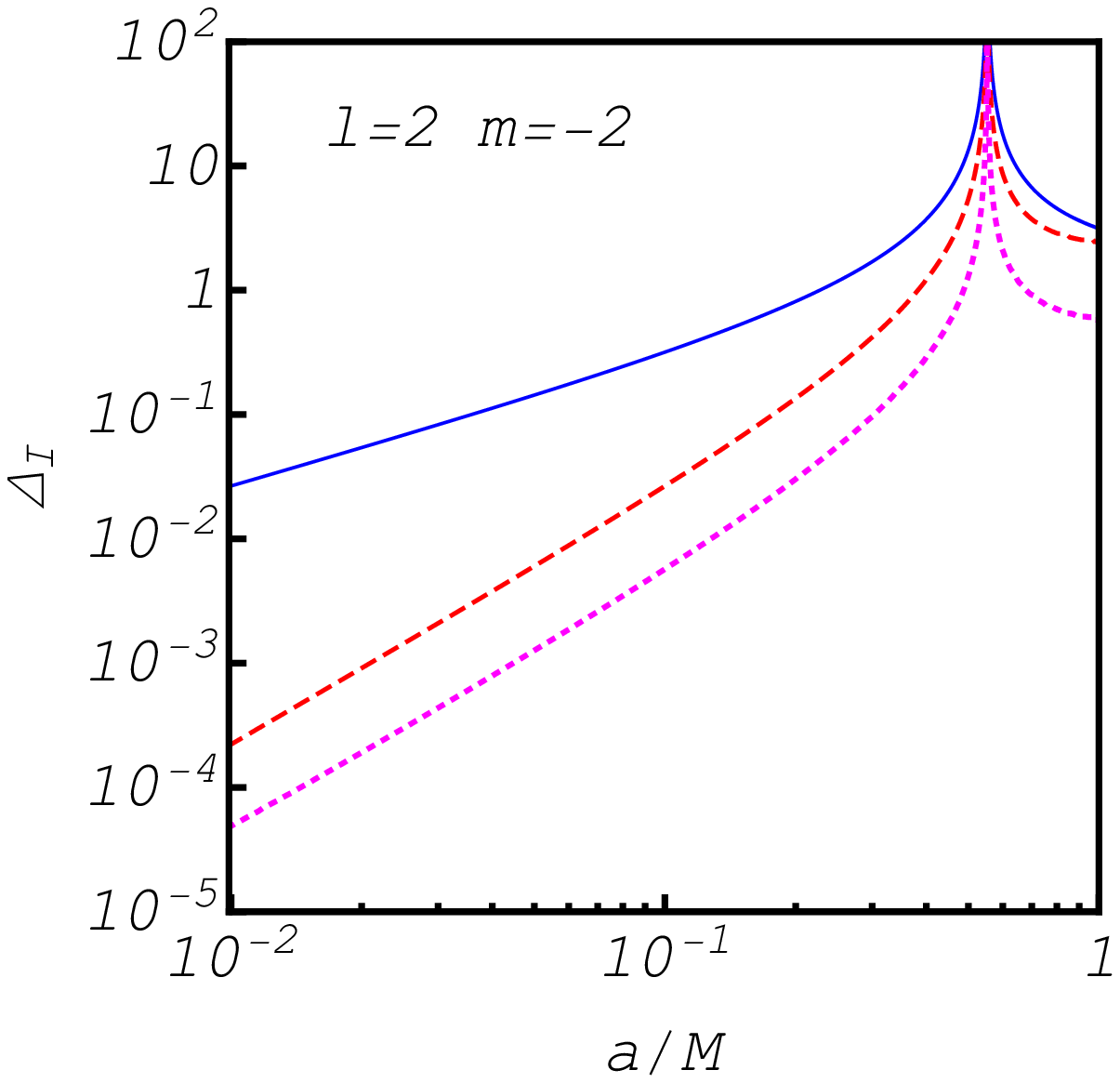}
\\
\includegraphics[width=4cm]{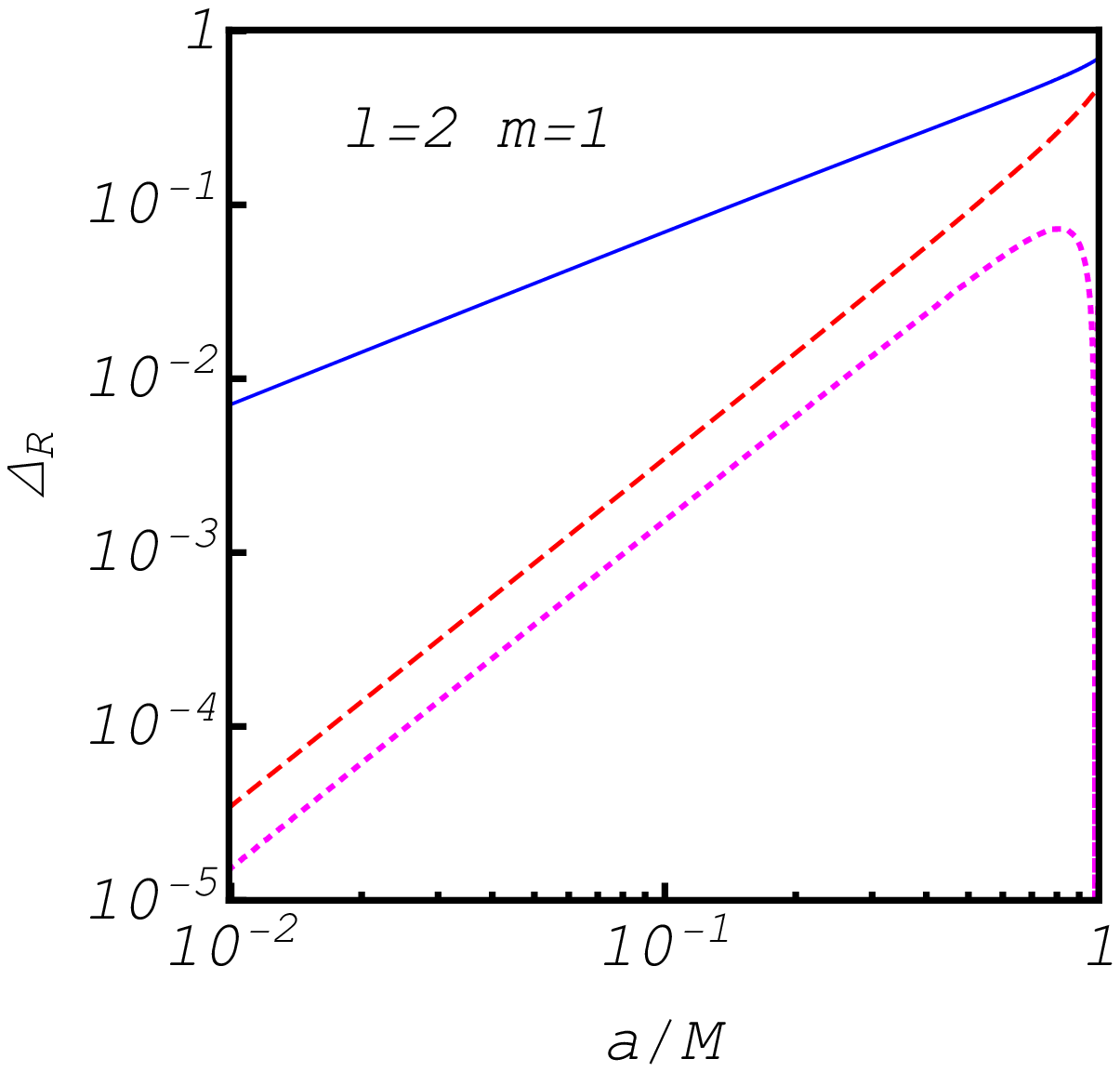}~~
\includegraphics[width=4cm]{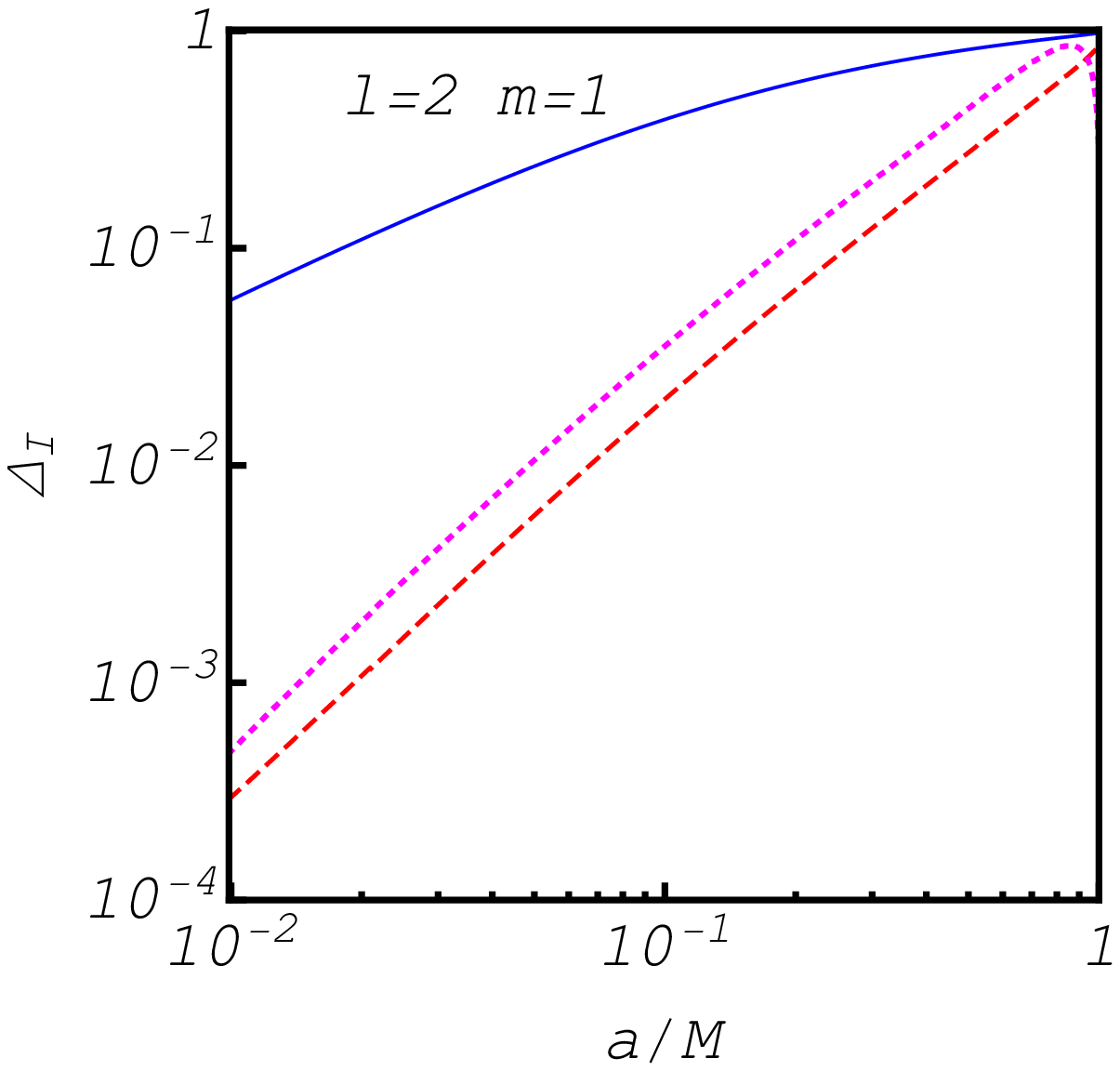}~~
\includegraphics[width=4cm]{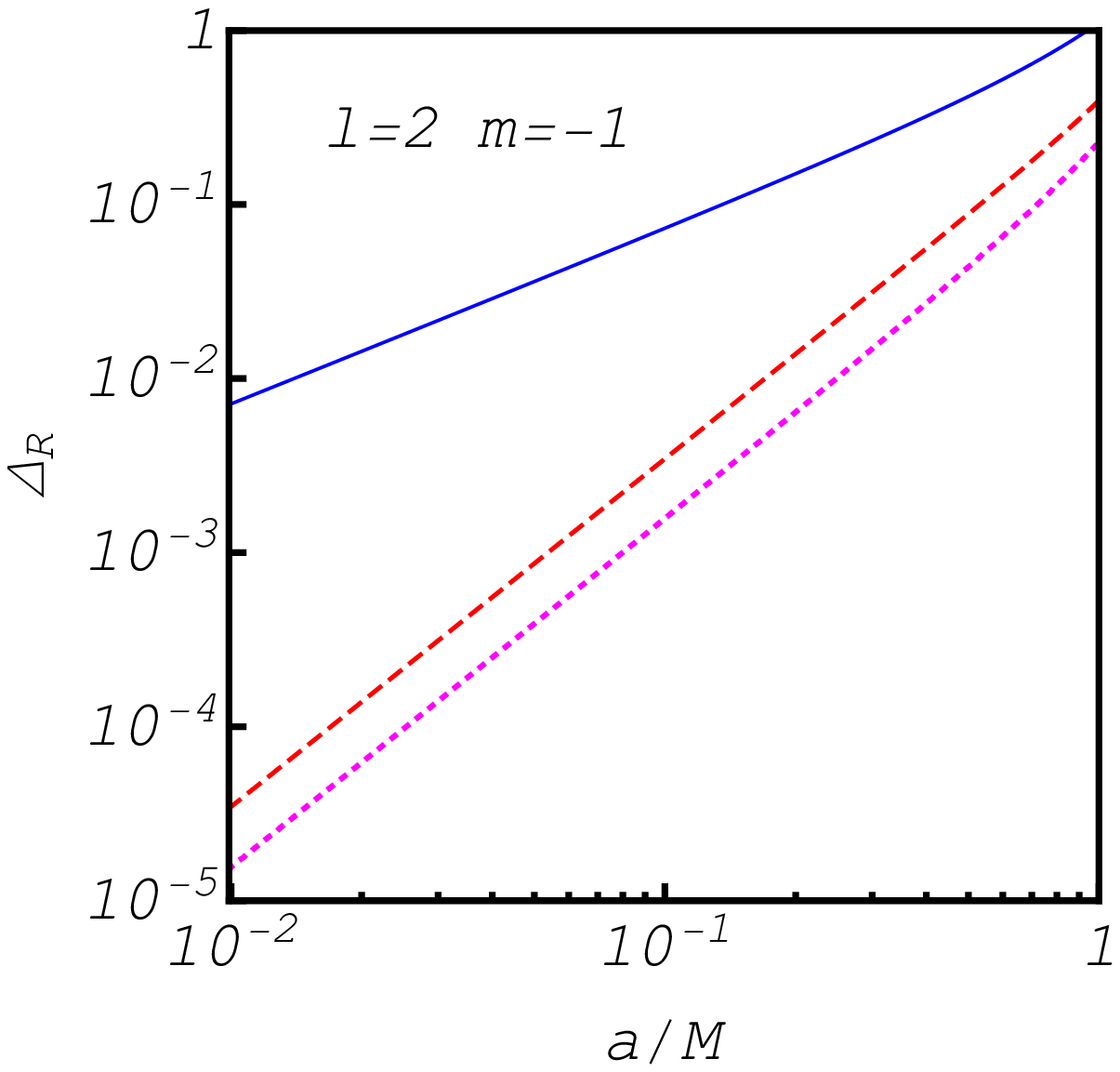}~~
\includegraphics[width=4cm]{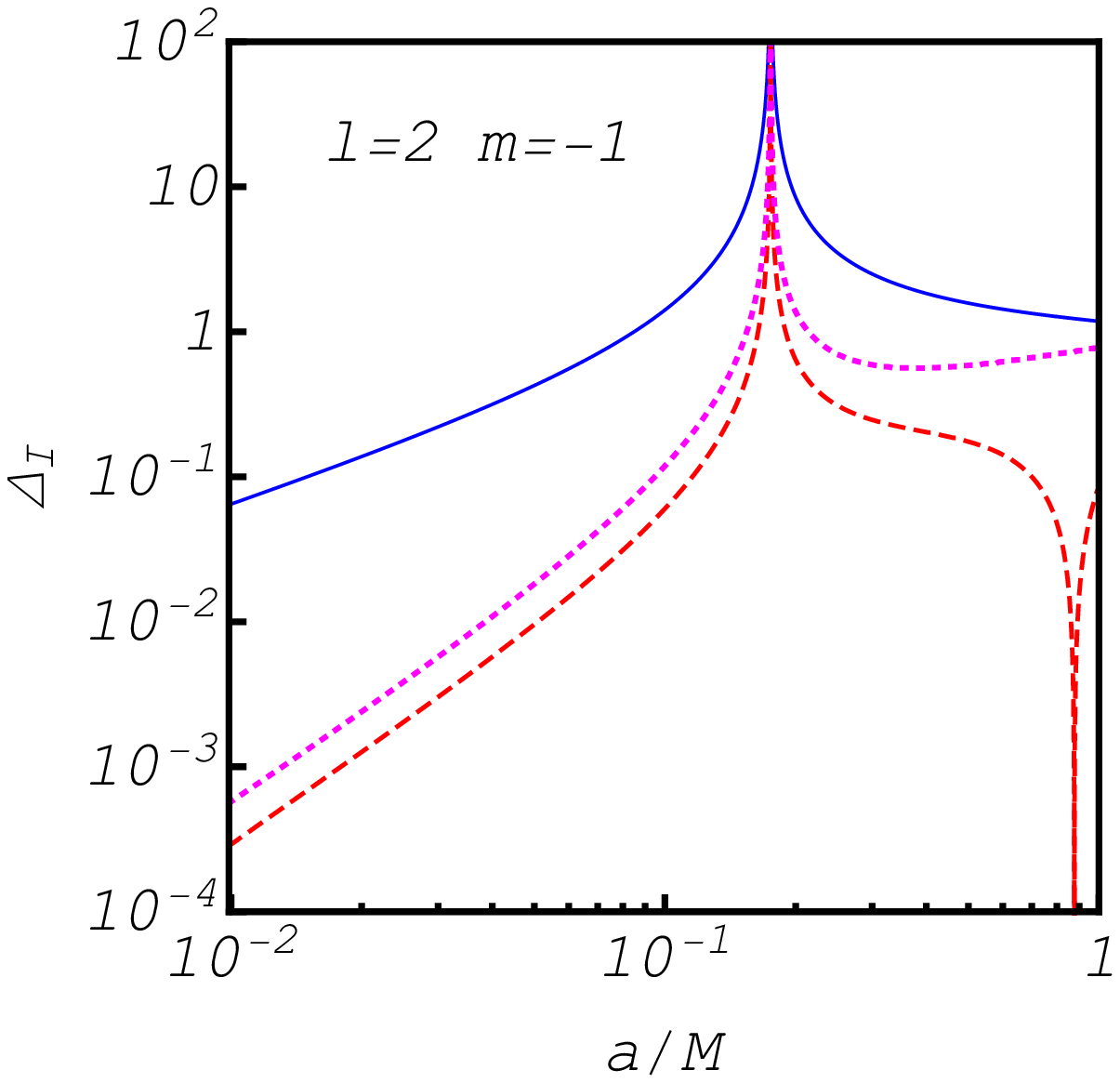}
\\
\includegraphics[width=4cm]{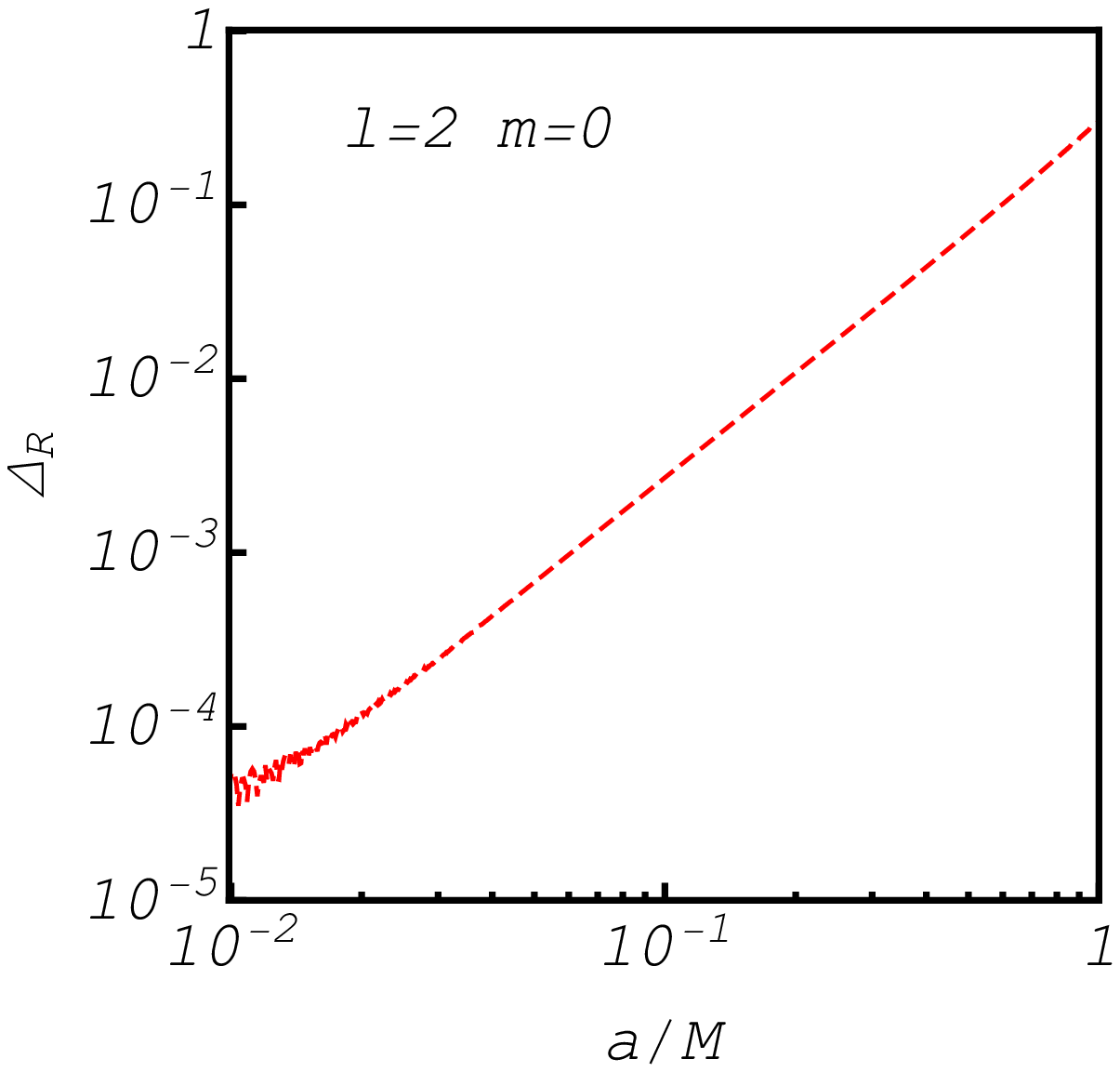}~~
\includegraphics[width=4cm]{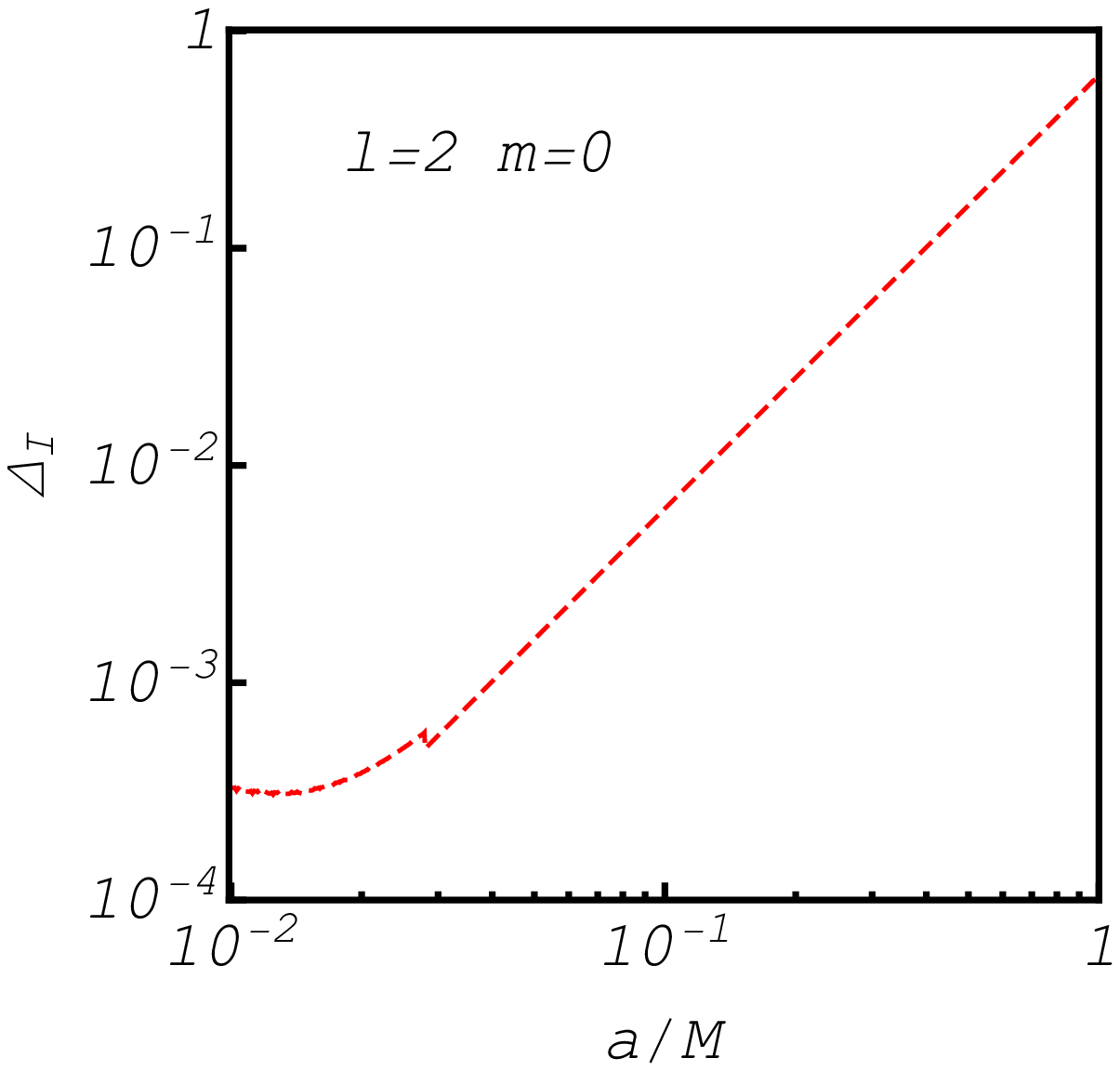}
\end{center}
\caption{\label{fig2} 
The error functions for $\ell = 2$ with $m = 0, \pm 1, \pm 2$.
Solid (blue), dashed (red), dotted (magenda) curves denote
the error functions for the first order expression, the second order expression and the Pad\'e approximant, respectively.
 }
\end{figure}

\begin{figure}
\begin{center}
\includegraphics[width=6cm]{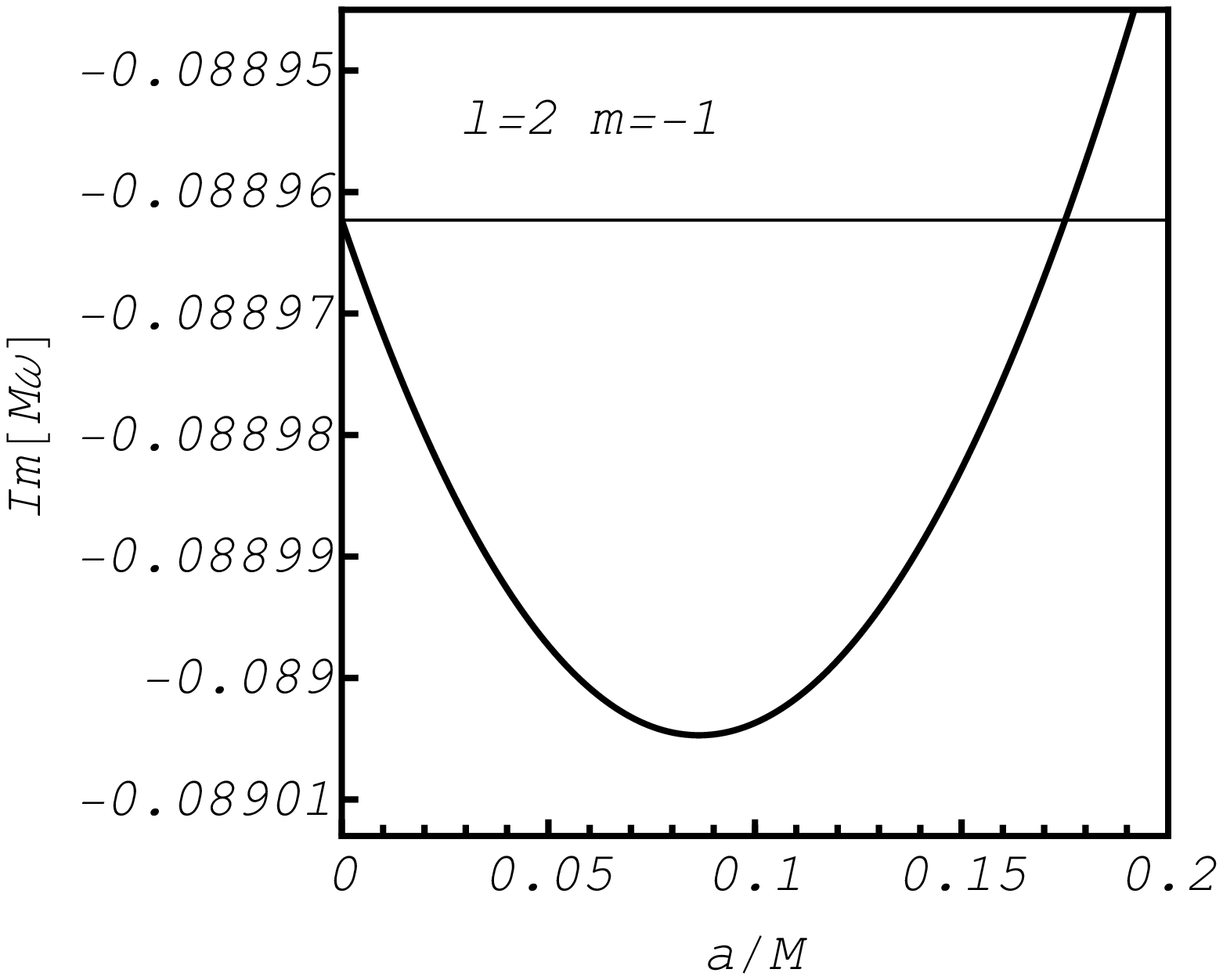}~~~
\includegraphics[width=6cm]{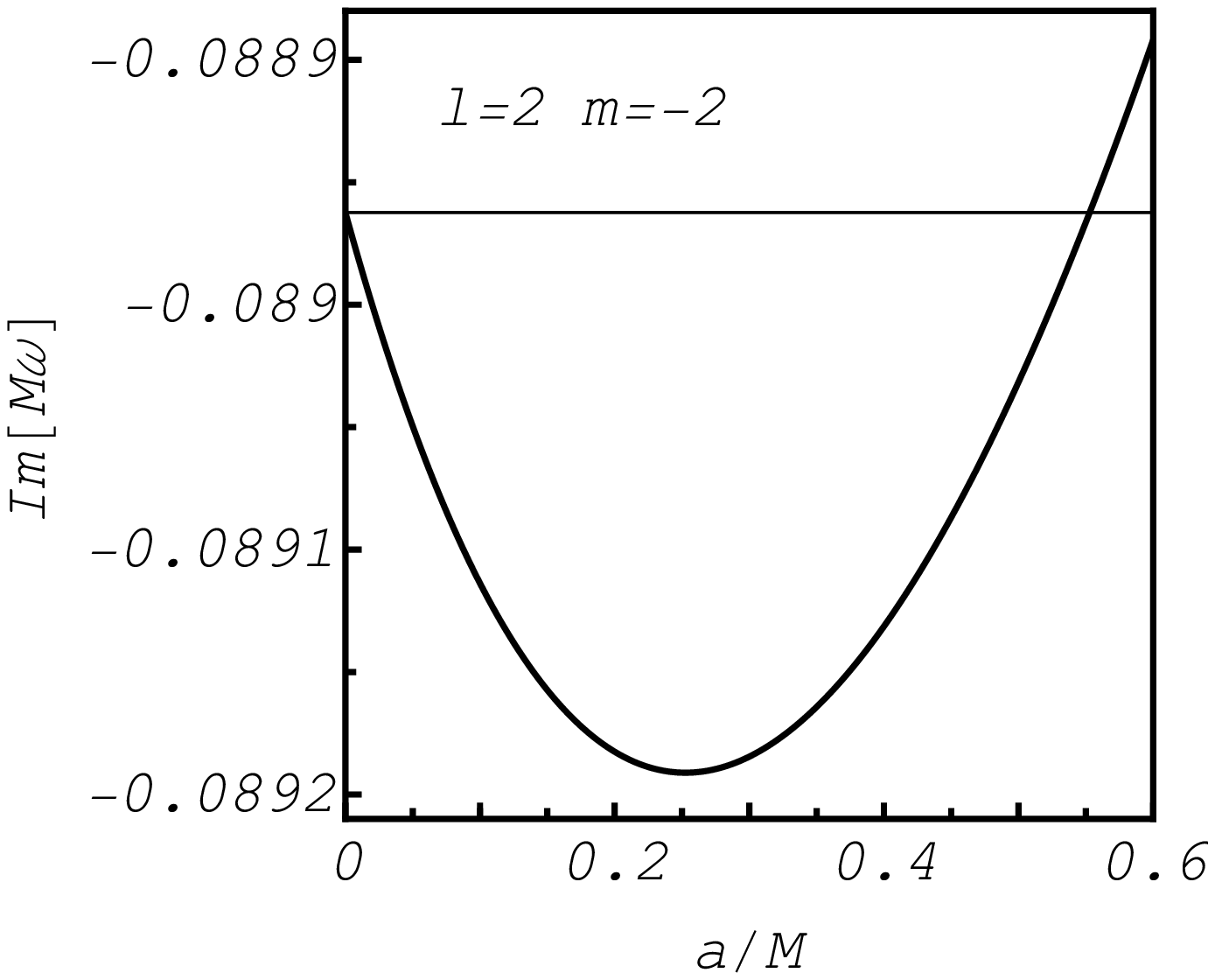}
\end{center}
\caption{\label{fig3}
The zoom plot for the imaginary part of the QNM frequencies for $m = -1, -2$ cases in Fig.~\ref{fig1}.
The values of the imaginary part coincide with those of the Schwarzschild case
at $a \simeq 0.175$ for $m = -1, \ell = 2$ and $a \simeq 0.553$ for $m=-2, \ell = 2$, respectively.
}
\end{figure}

In Fig.~\ref{fig2}, the error of our formula is plotted.
The error for the deviation from the Schwarzschild case is estimated by $\Delta_R$ and $\Delta_I$
\begin{align}
\Delta_R &= \left|\frac{{\rm Re}[(\omega_{\rm full}  - \omega_{\rm Sch})
 - (\omega^{\rm Kerr}_{\rm QNM}  - \omega_{\rm Sch})]}{{\rm Re}[\omega_{\rm full} - \omega_{\rm Sch}]}\right|,
\label{errorre}
\\
\Delta_I &= \left|\frac{{\rm Im}[(\omega_{\rm full}  - \omega_{\rm Sch})
 - (\omega^{\rm Kerr}_{\rm QNM}  - \omega_{\rm Sch})]}{{\rm Im}[\omega_{\rm full} - \omega_{\rm Sch}]}\right|,
\label{errorim}
\end{align}
where $\omega_{\rm Sch} = \Omega_0 M$, $\omega^{\rm Kerr}_{\rm QNM}$ is given by Eq.~\eqref{eq:kerrqnm2ndformula}
and $\omega_{\rm full}$ is a numerical result in~\cite{Berti:2005ys, Berti:2009kk, ringdowndata}.
We find that the error is very small for $a < 0.1$.
In Fig.~\ref{fig2},
the error function diverges at $a \simeq 0.175$ for $m = -1, \ell = 2$ and $a \simeq 0.553$ for $m=-2, \ell = 2$, respectively.
This is because that 
the denominator of Eq.~\eqref{errorim} accidentally vanishes at those points (see Fig.~\ref{fig3}).

In~\cite{Berti:2005ys}, fitting functions for the QNM frequencies of Kerr black hole from numerical calculation 
are discussed.
For $\ell = 2, m = 2$ fundamental modes, they are given by
\begin{align}
{\rm Re}[M\omega^{({\rm fit})}] &\simeq 1.5251 - 1.1568(1 - a/M)^{0.1292},
\notag\\&= 
0.3683 + 0.149459 a/M + 0.0650743 (a/M)^2 + O(a^3),
\\
\frac{{\rm Re}[\omega^{({\rm fit})}]}{2 |{\rm Im}[\omega^{({\rm fit})}]|} &\simeq
0.7000 + 1.4187 (1-a/M)^{-0.4990},
\notag\\&=
2.1187+0.707931 a/M +0.530595 (a/M)^2 +  O(a^3).
\end{align}
The same quantities from our result~\eqref{eq:kerrqnm2ndformula} become
\begin{align}
{\rm Re}[M\omega_{\rm [2nd]}^{\rm QNM}] &=
0.373672 + 0.125766  a/M  + 0.0717403(a/M)^2  +  O(a^3),
\\
\frac{{\rm Re}[\omega_{\rm [2nd]}^{\rm QNM}]}{2 |{\rm Im}[\omega_{\rm [2nd]}^{\rm QNM}]|} &= 
2.10017+0.753968 a/M +0.543279 (a/M)^2 + O(a^3).
\end{align}
While the fitting functions in~\cite{Berti:2005ys} are made from data in all regime $0 < a/M < 1$, 
they roughly reproduce our result Eq.~\eqref{eq:kerrqnm2ndformula} in the slow rotation regime.

\section{Summary and discussion}
In this paper, we derived semi-analytic expressions for quasinormal mode frequencies of slowly rotating Kerr black holes
up to the second order of the Kerr parameter $a$.
The key point is that 
the Chandrasekhar-Detweiler equation
(or the Sasaki-Nakamura equation) reduces to the Regge-Wheeler equation in the limit $a \to 0$,
and we can 
apply the parametrized black hole QNM ringdown formalism~\cite{Cardoso:2019mqo, McManus:2019ulj, Kimura:2020mrh}.
We also compared our result with the previous numerical calculations, and showed that
they agree very well at small Kerr parameter.
Our perturbative expression should provide useful information to improve fitting functions.

As a future work, it would be interesting to extend our result to the higher order Kerr parameters
because we have observed that the QNM frequencies at the quadratic order give a good approximation for not very small Kerr parameters.
If we have the higher order Kerr parameter extension, 
it is also possible to discuss radius of convergence of the series expansion and the singularity structure.
Another direction is an extension to modified gravities.
In modified gravities, master equations would be coupled systems, thus we expect that 
the parametrized black hole QNM ringdown formalism for coupled systems~\cite{McManus:2019ulj} will be useful.

\section*{Acknowledgments}
We would like to thank Norichika Sago and Takahiro Tanaka for useful comments on the paper.
The work of Y.H. is supported by JSPS KAKENHI Grant No. JP18K03657.
M.K. acknowledges support by MEXT Grant-in-Aid for Scientific Research on Innovative Areas 20H04746.
M.K. also thanks Theoretical Astrophysics Group at Kyoto University, where this work was initiated, for their hospitality.

\appendix

\section{QNM boundary condition}
\label{qnmbc}
The QNM boundary condition for the Chandrasekhar-Detweiler equation~\eqref{eq:cdeq} is given by~\cite{Detweiler:1977}
\begin{align}
X &\to e^{i\omega r_*} ~~{\rm for}~~r_* \to \infty,
\label{qbmbc1}
\\
X &\to e^{-ikr_*} ~~{\rm for}~~r_* \to -\infty,
\label{qbmbc2}
\end{align}
where $k = \omega - m a/(2 M r_+)$.
Near the event horizon, $k r_*$ behaves 
\begin{align}
k r_* \simeq \left(\omega - \frac{m}{r_+} \sqrt{\frac{r_-}{r_+}} + 2 \omega_0 \frac{r_-}{r_+}\right) r_+ \ln((r-r_+)/r_+)
+
r_+ \omega - m  \sqrt{\frac{r_-}{r_+}}  - r_+ \omega \frac{r_-}{r_+}
+
{\cal O}(r- r_+).
\end{align}
Thus, $e^{-ikr_*}$ approximately behaves 
\begin{align}
e^{-(\omega - (m/r_+)\sqrt{r_-/r_+} + 2 \omega_0 r_-/r_+ ) r_*^{\rm Sch}} \left({\rm const.} +
{\cal O}(r- r_+)\right),
\end{align}
where $r_*^{\rm Sch} = r + r_+ \ln((r-r_+)/r_+) \sim r_+ \ln((r-r_+)/r_+)$.

\section{Calculation from the Sasaki-Nakamura equation}
\label{sec:sasakinakamura}
\subsection{The Sasaki-Nakamura equation}
The Sasaki-Nakamura equation~\cite{Sasaki:1981kj, Sasaki:1981sx, Nakamura:1981kk} is given by
\begin{align}
\left(\frac{d^2}{dr_*^2} - F \frac{d}{dr_*} - U \right)X_{\rm SN} = 0,
\end{align}
where $r_*$ is same as Eq.~\eqref{eq:rstardef}.\footnote{
There is a relation between $X_{\rm SN}$ and the Teukolsky variable $R$ as
\begin{align}
R = \frac{1}{\eta} \left[
\left(\alpha + \frac{\partial_r\beta}{\Delta}\right) \chi  - \frac{\beta}{\Delta}\partial_r\chi
\right],
\end{align}
with $\chi = X_{\rm SN} \Delta/\sqrt{r^2 + a^2}$.
This also can be written as
\begin{align}
X_{\rm SN} = \sqrt{r^2 + a^2} r^2 \left[
\frac{d}{dr} - i \frac{K}{\Delta}
\right]\left[
\frac{d}{dr} - i \frac{K}{\Delta}
\right]
\left(\frac{R}{r^2}\right).
\end{align}
}
The functions $F$ and $U$ are given by
\begin{align}
F &= \frac{\partial_r \eta}{\eta} \frac{\Delta}{r^2 + a^2},
\\
\Delta &= r^2 - 2 M r + a^2,
\\
\eta &= c_0 + \frac{c_1}{r} + \frac{c_2}{r^2} + \frac{c_3}{r^3}
+
\frac{c_4}{r^4},
\\
c_0 &= - 12 i \omega M + \lambda (\lambda + 2) 
-12 a \omega (a\omega - m),
\\
c_1 &= 8i a (3 a \omega - \lambda(a\omega -m)),
\\
c_2 &= -24 i am (a \omega - m)
+12 a^2 (1-2 (a\omega -m)^2),
\\
c_3 &= 24 i a^3 (a\omega - m) - 24 M a^2,
\\
c_4 &= 12 a^4,
\end{align}
and
\begin{align}
U &= \frac{\Delta U_1}{(r^2 + a^2)^2} + G^2 + \frac{\Delta \partial_r G}{r^2 + a^2} - F G,
\\
G &= - \frac{2(r-M)}{r^2 + a^2} + \frac{r\Delta}{(r^2 + a^2)^2},
\\
U_1 &= V + \frac{\Delta^2}{\beta}
\left(
\partial_r \left(2 \alpha + \frac{\partial_r \beta}{\Delta}\right)
-\frac{\partial_r\eta}{\eta}
\left(
\alpha + \frac{\partial_r \beta}{\Delta}\right)\right),
\\
\alpha &= - i \frac{K \beta}{\Delta^2} + 3 i \partial_r K + \lambda + \frac{6\Delta}{r^2},
\\
\beta &= 2 \Delta \left(-i K + r - M - \frac{2\Delta }{r}\right) ,
\\
V &= - \frac{K^2 + 4 i (r-M)K}{\Delta} + 8 I \omega r +\lambda,
\\
K &= (r^2 + a^2)\omega - m a,
\end{align}
We note that the QNM boundary condition for the Sasaki-Nakamura equation is same as Eqs.~\eqref{qbmbc1} and~\eqref{qbmbc2}.

\subsection{First order expression}
Defining a new master variable $\tilde{X}_{\rm SN}$ as
\begin{align}
X_{\rm SN} &=: \tilde{X}_{\rm SN} \left(1 + Y_1 \sqrt{\frac{r_-}{r_+}}\right),
\\
Y_1 &= -\frac{4 m \sqrt{r_+ r_-}}{2 i \lambda_0 + i \lambda_0^2 + 6 r_+ \omega_0}
\left(\frac{\lambda_0}{r} + \frac{3 r_+}{2 r^2} \right),
\label{eqy1}
\end{align}
we can rewrite the Sasaki-Nakamura equation in the form
\begin{align}
& f \frac{d}{dr}
\left(f \frac{d}{dr}\right) \tilde{X}_{\rm SN}
+\left(\left(\omega - \frac{m}{r_+}\sqrt{\frac{r_-}{r_+}}\right)^2 
- f(V_{-} + \delta V 
)\right)\tilde{X}_{\rm SN} = 0,
\end{align}
with
\begin{align}
V_- &= \frac{\ell(\ell +1)}{r^2} - \frac{3r_+}{r^3},
\\
f &= 1- \frac{r_+}{r},
\\
\delta V &= \frac{1}{r_+^2}\sum_{j = 0}^5 \alpha_j^- \left(\frac{r_+}{r}\right)^j.
\end{align}
where $\alpha_j^- = \alpha_j^{({\rm 1st})}  \sqrt{r_-/r_+}$ and the explicit forms of $\alpha_j^{({\rm 1st})}$ are given by 
\begin{align}
\alpha_0^{({\rm 1st})} &= -2 m \omega_0 r_+,
\label{eqalpha0sn}
\\
\alpha_1^{({\rm 1st})} &= -2 m \omega_0 r_+ ,
\\
\alpha_2^{({\rm 1st})} &=  \frac{m r_+ \omega_0 (6 \lambda_1 r_+ \omega_0+i \lambda_0 ((\lambda_0+2) \lambda_1+8))}{6 r_+ \omega_0+i \lambda_0 (\lambda_0+2)},
\\
\alpha_3^{({\rm 1st})} &= \frac{8 m (\lambda_0+3 i r_+ \omega_0)}{6 r_+ \omega_0+i \lambda_0 (\lambda_0+2)} ,
\\
\alpha_4^{({\rm 1st})} &=  -\frac{12 (\lambda_0-5) m}{6 r_+ \omega_0+i \lambda_0 (\lambda_0+2)},
\\
\alpha_5^{({\rm 1st})} &= -\frac{72 m}{6 r_+ \omega_0+i \lambda_0 (\lambda_0+2)},
\label{eqalpha5sn}
\end{align}
where $\omega_0 = 2\Omega_0/r_+$.
The QNM frequency for Kerr black hole for small $r_-$ becomes
\begin{align}
\omega_{\rm QNM}^{\rm Kerr}  = \frac{2\Omega_0}{r_+} + \sum_{j = 0}^5 \alpha_j^- e_j^- +  \frac{m}{r_+}\sqrt{\frac{r_-}{r_+}} + {\cal O}(r_-).
\end{align}
Apparently, this result looks different from that for the Chandrasekhar-Detweiler equation.
However in fact, they become same equations if we use the recursion relation for $e_j^-$ in~\cite{Kimura:2020mrh}.

\subsection{Second order expression}
In the second order case, we introduce a variable $\tilde{X}_{\rm SN}$ as
\begin{align}
X_{\rm SN} &=: \tilde{X}_{\rm SN} \left(1 + Y_1 \sqrt{\frac{r_-}{r_+}} + Y_2 \frac{r_-}{r_+}\right),
\end{align}
where $Y_1$ is same as Eq.~\eqref{eqy1} and $Y_2$ is given by
\begin{align}
Y_2 &= 
-\frac{r_+}{2 r^4 (\lambda_0^2+2 \lambda_0-6 i (r_+ \omega_0))^2}
\Big[
24 m r^2 r_+ \omega_1 (2 \lambda_0 r+3 r_+)
\notag\\&
+m^2 
\Big(
8 i r^3 r_+ \omega_0 (\lambda_0^2 \lambda_1+12 \lambda_0+6 i \lambda_1 r_+ \omega_0)
\notag\\&
+8 r^2 r_+ (\lambda_0^2+\lambda_0 (6+3 i \lambda_1 r_+ \omega_0)+3 i \lambda_1 r_+ \omega_0) -48 \lambda_0 r r_+^2-36 r_+^3\Big)
\notag\\&
-r (\lambda_0^2+2 \lambda_0-6 i r_+ \omega_0) 
\Big(
r^2 (\lambda_0^2+\lambda_0 (2-8 i r_+ \omega_0)+18 i r_+ \omega_0)
\notag\\&
+r r_+ (\lambda_0^2+2 \lambda_0-18 i r_+ \omega_0+12)-12 r_+^2
\Big)
\Big].
\end{align}
Then, the Sasaki-Nakamura equation becomes
\begin{align}
& f \frac{d}{dr}
\left(f \frac{d}{dr}\right) \tilde{X}_{\rm SN}
+\left(\left(\omega - \frac{m}{r_+}\sqrt{\frac{r_-}{r_+}} 
+
 2 \omega_0 \frac{r_-}{r_+}
\right)^2 
- f (V_{-} + \delta V)\right)\tilde{X}_{\rm SN} = 0,
\end{align}
with
\begin{align}
\delta V &= \frac{1}{r_+^2}\sum_{j = 0}^7 \alpha_j^- \left(\frac{r_+}{r}\right)^j.
\end{align}
where
$\alpha_j^- = \alpha_j^{({\rm 1st})} \sqrt{r_-/r_+} + \alpha_j^{({\rm 2nd})} r_-/r_+$.
The first order corrections $\alpha_j^{({\rm 1st})}$ 
are same as Eqs.~\eqref{eqalpha0sn}-\eqref{eqalpha5sn} (but $\alpha_6^{({\rm 1st})} = \alpha_7^{({\rm 1st})} = 0$),
and the second order corrections $\alpha_j^{({\rm 2nd})}$ are given by 
\begin{align}
\alpha_0^{({\rm 2nd})} &= m^2-2 m \omega_1 r_+ +4 (\omega_0 r_+)^2,
\\
\alpha_1^{({\rm 2nd})} &= m^2-2 m \omega_1 r_+ +2 (\omega_0 r_+)^2,
\\
\alpha_2^{({\rm 2nd})} &= -\frac{8 (\lambda_0-3) (\omega_0 r_+)^2}{\lambda_0 (\lambda_0+2)-6 i \omega_0 r_+ }
+\lambda_2 (\omega_0 r_+)^2
\notag\\&
+\frac{m \omega_1 r_+  ((\lambda_0+2) \lambda_0^2 ((\lambda_0+2) \lambda_1+8)-12 i (\lambda_0+2) \lambda_0 \lambda_1 \omega_0 r_+ -36 \lambda_1 (\omega_0 r_+)^2)}{(\lambda_0 (\lambda_0+2)-6 i \omega_0 r_+ )^2}
\notag\\& 
+\frac{m^2 (\lambda_0^2 (\lambda_0+2)^2-4 (\omega_0 r_+)^2 (2 \lambda_0 (\lambda_0 \lambda_1+12)+9)-12 i \lambda_0 (\lambda_0+2) \omega_0 r_+ -48 i \lambda_1 (\omega_0 r_+)^3 )}{(\lambda_0 (\lambda_0+2)-6 i \omega_0 r_+ )^2},
\\
\alpha_3^{({\rm 2nd})} &= 
\frac{\lambda_0^3+2 i \lambda_0 (\omega_0 r_+ +2 i)-12 \omega_0 r_+  (2 \omega_0 r_+ +3 i)}{\lambda_0 (\lambda_0+2)-6 i \omega_0 r_+ }
+\frac{24 \lambda_0 (\lambda_0+4) m \omega_1 r_+ }{(\lambda_0 (\lambda_0+2)-6 i \omega_0 r_+ )^2}
\notag\\& 
+\frac{m^2 (\lambda_0^2 (\lambda_0+2)^2-12 (\omega_0 r_+)^2 (4 (\lambda_0+2) \lambda_1+3)+4 i \lambda_0 \omega_0 r_+  (\lambda_0 (2 \lambda_1-7)+42))}{(\lambda_0 (\lambda_0+2)-6 i \omega_0 r_+ )^2},
\\
\alpha_4^{({\rm 2nd})} &=
\frac{3 (\lambda_0 (3 \lambda_0-8 i \omega_0 r_+ -2)+70 i \omega_0 r_+ -40)}{2 \lambda_0 (\lambda_0+2)-12 i \omega_0 r_+ }
-\frac{72 (\lambda_0-5) m \omega_1 r_+ }{(\lambda_0 (\lambda_0+2)-6 i \omega_0 r_+ )^2}
\notag\\& 
+\frac{m^2 (-12 i \omega_0 r_+  (((\lambda_0-10) \lambda_0-10) \lambda_1+32 \lambda_0)-8 (\lambda_0-5) \lambda_0 (\lambda_0+6)+72 \lambda_1 (\omega_0 r_+)^2)}{(\lambda_0 (\lambda_0+2)-6 i \omega_0 r_+ )^2},
\\
\alpha_5^{({\rm 2nd})} &=
\frac{12 (\lambda_0-6 i \omega_0 r_+ +15)}{\lambda_0 (\lambda_0+2)-6 i \omega_0 r_+ }
-\frac{432 m \omega_1 r_+ }{(\lambda_0 (\lambda_0+2)-6 i \omega_0 r_+ )^2}
\notag\\& 
+\frac{m^2 (-24 \lambda_0 (\lambda_0+50)-144 i \omega_0 r_+  (\lambda_0 \lambda_1+\lambda_1+1))}{(\lambda_0 (\lambda_0+2)-6 i \omega_0 r_+ )^2},
\\
\alpha_6^{({\rm 2nd})} &= 
-\frac{126}{\lambda_0 (\lambda_0+2)-6 i \omega_0 r_+ } + \frac{144 (7 \lambda_0-8) m^2}{(\lambda_0 (\lambda_0+2)-6 i \omega_0 r_+ )^2},
\\
\alpha_7^{({\rm 2nd})} &= \frac{1296 m^2}{(\lambda_0 (\lambda_0+2)-6 i \omega_0 r_+ )^2},
\end{align}
where $\omega_1$ is given by
\begin{align}
\omega_1  = 
\sum_{j = 0}^5 \alpha_j^{({\rm 1st})}  e_j^- +  \frac{m}{r_+}.
\end{align}
Then, the QNM frequency becomes
\begin{align}
\omega_{\rm QNM}^{\rm Kerr}  = \frac{2\Omega_0}{r_+}
+ \sum_{j = 0}^7 \alpha_j^{-} e_j^- 
+ \sum_{j,k = 0}^5 \alpha_j^{-} \alpha_k^{-} e_{jk}^- 
+ \frac{m}{r_+}\sqrt{\frac{r_-}{r_+}}
 - 2\omega_0 \frac{r_-}{r_+}+ {\cal O}(r_-^{3/2}).
\end{align}
We note that this equation reproduces the same numerical value as Eq.~\eqref{eq:kerrqnm2ndformula}.

\end{document}